\documentclass[12pt]{article}
\pdfoutput=1 
\usepackage{multicol}
\usepackage{cancel,slashed}
\usepackage{bbm}
\usepackage{amsfonts}
\usepackage[usenames,dvipsnames]{xcolor}
\usepackage{amsmath}
\usepackage{amssymb}
\usepackage{graphicx}
\usepackage{subcaption}
\usepackage{cite}
\usepackage{hyperref}
\usepackage{bm}

\usepackage{latexsym}
\usepackage{color}
\usepackage{cancel,slashed}
\usepackage{tikz}

\makeatletter
\def\@xfootnote[#1]{%
  \protected@xdef\@thefnmark{#1}%
  \@footnotemark\@footnotetext}
\newcommand{\bmat}{\left(\begin{array}}
\newcommand{\emat}{\end{array}\right)}

\def\b{\beta}
\def\g{\gamma}

\def\-{\hphantom{-}}

\def\s2{\frac{1}{\sqrt2}}

\def\beq{\begin{equation}}
\def\eeq{\end{equation}}
\def\beqa{\begin{eqnarray}}
\def\eeqa{\end{eqnarray}}

\def\re{{\rm Re \,}}

\def\Dsl{\,\raise.15ex\hbox{/}\mkern-13.5mu D} %this one can be subscripted

\def\re{\mbox{Re}}

\def\be{\begin{equation}}
\def\ee{\end{equation}}
\def\bea{\begin{eqnarray}}
\def\eea{\end{eqnarray}}
\def\raw{\rightarrow}

\def\IN{\mathbb{N}}
\def\IZ{\mathbb{Z}}
\def\IR{\mathbb{R}}

\def\b{{\beta}}

\def\eps{{\epsilon}}

\def\sig{{\sigma}}

\def\g{{\gamma}}

\makeatletter
\newsavebox{\@brx}
\newcommand{\llangle}[1][]{\savebox{\@brx}{\(\m@th{#1\langle}\)}%
  \mathopen{\copy\@brx\kern-0.5\wd\@brx\usebox{\@brx}}}
\newcommand{\rrangle}[1][]{\savebox{\@brx}{\(\m@th{#1\rangle}\)}%
  \mathclose{\copy\@brx\kern-0.5\wd\@brx\usebox{\@brx}}}
\makeatother

\def\sm2{{\mbox{\small 2}}}

\newcommand{\bp}{\begin{pmatrix*}[r]}  
\newcommand{\ep}{\end{pmatrix*}}  
\newcommand{\bpp}{\begin{pmatrix}}  
\newcommand{\epp}{\end{pmatrix}}  
\newcommand{\bcd}{\begin{center}
\begin{tikzcd}}
\newcommand{\ecd}{\end{tikzcd} \end{center}}

\def\1{\mathbb{1}}

\newenvironment{eqn}{\begin{equation}\begin{aligned}}{\end{aligned}\end{equation}\noindent}
\newenvironment{eqn*}{\begin{equation*}\begin{aligned}}{\end{aligned}\end{equation*}\noindent}

\definecolor{myblue}{RGB}{0,79,219}
\definecolor{myred}{RGB}{255,85,85}

\begin{document}
\pagestyle{plain}

%----------------------------------------------------------------------%
%  numbering equations with section number
%----------------------------------------------------------------------%
\makeatletter
\@addtoreset{equation}{section}
\makeatother
\renewcommand{\theequation}{\thesection.\arabic{equation}}
%----------------------------------------------------------------------%
%  title page
%----------------------------------------------------------------------%
\pagestyle{empty}
%\vspace*{1.0in}
\rightline{ IFT-UAM/CSIC-19-161}
%\rightline{\tt hep-th/yymmnnn}
\vspace{0.5cm}
\begin{center}
{
	\Huge{{Instanton Corrections and Emergent Strings}}
\\[15mm]}
\normalsize{Florent Baume,$^1$ Fernando Marchesano,$^1$ and Max Wiesner$^{1,2}$\\[10mm]}
\small{
${}^1$Instituto de F\'{\i}sica Te\'orica UAM-CSIC, Cantoblanco, 28049 Madrid, Spain \\[2mm] 
${}^2$ Departamento de F\'{\i}sica Te\'orica, 
Universidad Aut\'onoma de Madrid, %Cantoblanco, 
28049 Madrid, Spain
\\[8mm]} 
\small{\bf Abstract} \\[5mm]
\end{center}
\begin{center}
\begin{minipage}[h]{15.0cm} 

We study limits of infinite distance in the moduli space of 4d ${\cal N}=2$ string compactifications, in which instanton effects dominate. We first consider trajectories in the hypermultiplet moduli space of type IIB Calabi-Yau compactifications. We observe a correspondence between towers of D-brane instantons and D-brane 4d strings, such that the lighter the string the more relevant the instanton effects are. The dominant instantons modify the classical trajectory such that the lightest D-brane string  becomes tensionless even faster, while the other strings are prevented to go below the fundamental string tension. This lightest string is dual to a fundamental type IIB string and realises the Emergent String Conjecture. We also consider the vector multiplet moduli space of type I string theory on $K3 \times T^2$, where quantum corrections can also become significant. Naively, we only find trajectories that correspond to decompactification limits, in apparent contradiction with the picture obtained in some dual setup.

\end{minipage}
\end{center}
\newpage
%----------------------------------------------------------------------%
%  Resetting of counters
%----------------------------------------------------------------------%
\setcounter{page}{1}
\pagestyle{plain}
\renewcommand{\thefootnote}{\arabic{footnote}}
\setcounter{footnote}{0}
%----------------------------------------------------------------------%
%  Paper begins
%----------------------------------------------------------------------%

%\end{document}

\tableofcontents

%\newpage

%%%%%%%%%%%%%%%%%%%%%%%%%%%%%%%%%%%%%%%%%%%%%%%%%
%%%%%%%%%%%%%%%%%%%%%%%%%%%%%%%%%%%%%%%%%%%%%%%%%
%%%%%%%%%%%%%%%%%%%%%%%%%%%%%%%%%%%%%%%%%%%%%%%%%
\section{Introduction}\label{s:intro}
%%%%%%%%%%%%%%%%%%%%%%%%%%%%%%%%%%%%%%%%%%%%%%%%%
%%%%%%%%%%%%%%%%%%%%%%%%%%%%%%%%%%%%%%%%%%%%%%%%%
%%%%%%%%%%%%%%%%%%%%%%%%%%%%%%%%%%%%%%%%%%%%%%%%%
Recently, the swampland programme has stirred the field of string phenomenology, 
by aiming at drawing a clear line between effective field theories
(EFTs) that have a consistent UV completion including quantum gravity from
those that do not. While the landscape of string compactifications is expected
to either cover or provide a representative subset of the former set of
theories, the latter set is dubbed to form the \textit{swampland}
\cite{Vafa:2005ui}.

To differentiate between what belongs to the swampland and what does not, a
series of swampland conjectures has been proposed, see \cite{Brennan:2017rbf,
Palti:2019pca} for recent reviews. In this context, the Swampland Distance
Conjecture (SDC) \cite{Ooguri:2006in} has played a central r\^ole, not only
because it describes inherent limitations of EFTs at long distances in field
space, but also because non-trivial connections between several swampland
conjectures have been drawn through its study. The SDC has been tested in
significant detail for string compactifications with eight supercharges
\cite{Palti:2017elp,Grimm:2018ohb,Heidenreich:2018kpg,Blumenhagen:2018nts,
Lee:2018urn,Lee:2018spm,Grimm:2018cpv,Buratti:2018xjt,Corvilain:2018lgw,Joshi:2019nzi,Marchesano:2019ifh,
Lee:2019xtm, Grimm:2019wtx, Erkinger:2019umg, Lee:2019wij}, where there is an
actual moduli space, and also analysed in the context of four supercharges
\cite{Baume:2016psm,Klaewer:2016kiy,Valenzuela:2016yny,Blumenhagen:2017cxt,Hebecker:2017lxm,Landete:2018kqf,Ooguri:2018wrx,Hebecker:2018fln,Gonzalo:2018guu,Lee:2019tst,Blumenhagen:2019qcg,Font:2019cxq},
where the field space is endowed with a potential. 

As a by-product of these analyses, the notion of emergence has been
established, stating that the infinite distances occurring at boundaries of
moduli spaces are in fact a consequence of illegally integrating out the states
that become massless in this limit
\cite{Harlow:2015lma,Heidenreich:2017sim,Grimm:2018ohb,Heidenreich:2018kpg,Palti:2019pca}.
Recently, this notion has been refined in terms of the Emergent String
Conjecture \cite{Lee:2019wij}, which states that in fact at all
equi-dimensional infinite distance limits, the theory reduces to an
asymptotically tensionless, weakly-coupled string theory. As such, the tower of
light modes required by the SDC are given by the oscillations of a critical
heterotic or Type II fundamental string. The refinement of specialising to
equi-dimensional limits excludes trajectories of infinite distance
that are partial decompactification limits due to the appearance of a KK-like
spectrum dominating the EFT breakdown. 

In \cite{Lee:2019wij} the Emergent String Conjecture was tested in the K\"ahler
moduli space of Calabi--Yau compactifications of M-theory to 5d and type IIA to
4d. The latter describes the vector multiplet moduli space of type II
Calabi--Yau compactifications, on which the masses of BPS particles depend and
whose metric is known exactly. These two features provide powerful tools to
classify the different infinite distance trajectories, and in particular
determine whether they are equi-dimensional or not. 

The purpose of this work is to further test the Emergent String Conjecture,
considering cases where instanton corrections significantly modify the moduli
space metric. In particular we consider infinite distance trajectories in the
hypermultiplet space of type IIB Calabi--Yau compactifications to 4d, making
use of the analysis performed in \cite{Marchesano:2019ifh}. This moduli space
contains the Calabi--Yau K\"ahler moduli and so, classically, looks very
similar to a type IIA CY vector multiplet moduli space. Nevertheless, there are
two important differences between these two cases that directly affect the
analysis of infinite distance trajectories. First, the masses of charged, BPS
particles of the compactification do not depend on the hypermultiplets. As a
result, when going to the boundary of the HM moduli space the only candidates
to drive a decompactification limit are standard Kaluza--Klein modes, in stark
contrast with the plethora of possibilities observed in \cite{Lee:2019wij}.
Instead, the hypermultiplets control the action of 4d instantons, which non-trivially
correct the classical metric. As pointed out in \cite{Marchesano:2019ifh} (see
also \cite{Grimm:2019wtx}), the trajectories in vector moduli space where
towers of particles become asymptotically massless correspond, via the c-map,
to hypermultiplet trajectories in which towers of towers of instantons develop
a vanishing action. In this regime the instanton effects become very large and
dominate over the classical contribution to the hypermultiplet moduli space
metric, again in dramatic contrast to the vector multiplet case.

It follows from these remarks that CY hypermultiplet spaces provide a very
interesting arena to test the Emergent String Conjecture. On the one hand,
classically one may apply the results of \cite{Grimm:2018ohb,Lee:2019wij} to
classify potential infinite distance limits. On the other hand, the physics in
such limits will be very different from their vector multiplet counterpart,
because {\it i)} there are less candidates to drive a decompactification limit
and {\it ii)} instantons will significantly modify the EFT when proceeding
along such trajectories. In particular, as we will see they will modify the
tensions of the 4d strings, which are the second key ingredient of the Emergent
String Conjecture. 

As a matter of fact, considering the spectrum of 4d strings that are sensitive
to the hypermultiplet sector gives a new perspective on the whole analysis.
Note that for each 4d string made up from a $D(p+2)$-brane wrapping an internal
$(p+1)$-cycle of $X$ there will be a D-brane instanton tower constructed by
wrapping $m$ D$p$-branes on  the same cycle. A limit in which the instanton action
asymptotically vanishes corresponds to a similar behaviour for the string
tension. As a result, in those regions of the hypermultiplet space in which
some strings become light certain instanton sectors become very relevant. This
poses an interesting challenge for the Emergent String Conjecture, which states
that there should be a unique string dominating the EFT breakdown. In the
infinite distance trajectories which we analyse several strings become
asymptotically tensionless, but there is one which is the lightest, and the
corresponding instanton corrections dominate over the classical action and the
remaining corrections. It turns out that the effect of such corrections is to
boost the speed at which the lightest string becomes tensionless, by bending
the trajectory and making it move along the 4d dilaton direction. Other 4d
strings also tend to a tensionless limit along the trajectory, but at a slower,
universal rate which makes them irrelevant for the EFT breakdown. Finally,
quantum corrections are such that from a certain point on the trajectory the
scale of the lightest string coincides with the scale of a KK tower of states,
similarly to the equi-dimensional limits analysed in \cite{Lee:2019wij}. In
this way the lightest D-brane 4d string along the trajectory, dual to a
fundamental IIB string, is able to realise the Emergent String Conjecture. 

We apply a similar analysis to the vector multiplet moduli space of type I
string theory on $K3\times T^2$. This compactification yields a 4d ${\cal N}=2$
EFT, but the vector multiplet sector contains both K\"ahler and complex structure moduli, 
and it is subject to several quantum corrections. Therefore its infinite distance 
limits are not obviously captured by
the classifications made in \cite{Grimm:2018ohb,Lee:2019wij}. We consider
several limits in this setup, taking into account the quantum corrections to
the K\"ahler potential  computed in \cite{Camara:2008zk}. In all such 
 limits we observe plain decompactification limits.
When compared to the analysis performed in dual setups this seems to 
raise a discrepancy for one of them. 

The rest of the paper is structured as follows. In Section \ref{sec:setup}, we
describe the type IIB hypermultiplet moduli space and we introduce the
different classical infinite distance trajectories that we later analyse in
Section \ref{sec:QuantumCorrections}. In there, we elaborate on the
correspondence between 4d string tensions and quantum corrections and use it as
an organising principle to see how at each of the trajectories endpoints a
weakly-coupled fundamental type II string emerges. In Section \ref{sec:typeI}
we analyse infinite distance trajectories in the vector multiplet sector of
type I strings on $K3\times T^2$, obtaining always decompactification limits.
For one of them, we point out a mismatch with results obtained in a dual setup.
We draw our conclusions in section \ref{sec:conclu}.

Several technical details have been relegated to the appendices. Appendix
\ref{app:PlanckMass} keeps track of the relations between the 4d Planck mass
and the string scale in the different frames used in the main text. Appendix
\ref{app:metric} collects all the necessary expressions for the quantum
corrected metric in hypermultiplet moduli space. Appendix
\ref{app:D3-instantons} reviews how to define a contact potential for mutually
local D-brane instantons including D3-branes. 

%%%%%%%%%%%%%%%%%%%%%%%%%%%%%%%%%%%%%%%%%%%%%%%%%
%%%%%%%%%%%%%%%%%%%%%%%%%%%%%%%%%%%%%%%%%%%%%%%%%
%%%%%%%%%%%%%%%%%%%%%%%%%%%%%%%%%%%%%%%%%%%%%%%%%
\section{Limits in Type IIB Hypermultiplet Moduli Space}\label{sec:setup}
%%%%%%%%%%%%%%%%%%%%%%%%%%%%%%%%%%%%%%%%%%%%%%%%%
%%%%%%%%%%%%%%%%%%%%%%%%%%%%%%%%%%%%%%%%%%%%%%%%%
%%%%%%%%%%%%%%%%%%%%%%%%%%%%%%%%%%%%%%%%%%%%%%%%%

Let us consider type IIB string theory compactified on a Calabi-Yau threefold $X$. The low energy four-dimensional EFT exhibits $\mathcal{N}=2$ supersymmetry with a moduli space $\mathcal{M}$ that factorises in hyper- and vector multiplet sectors:
\begin{align}
      \mathcal{M} = \mathcal{M}_\text{VM} \times \mathcal{M}_\text{HM}\,. 
\end{align}
On the one hand, the vector multiplet moduli sector $\mathcal{M}_\text{VM}$ is
a $h^{2,1}$-dimensional special K\"ahler manifold.  On the other hand, each
hypermultiplet transforms in the fundamental representation of the $SU(2)$
R-symmetry group, endowing $\mathcal{M}_{HM}$ with a quaternionic-K\"ahler
structure. While the possible infinite distance limits in
$\mathcal{M}_\text{VM}$ can be classified via mixed Hodge structure methods
\cite{Grimm:2018cpv,Grimm:2018ohb,Grimm:2019bey}, the hypermultiplet sector has
been comparatively largely unexplored in the context of the Swampland Distance
Conjecture, except for \cite{Marchesano:2019ifh} (see also \cite{Grimm:2019wtx}
for a study of the type IIA mirror dual). In this section and the next, this
part of the moduli space will be the main focus of our study.

In a type IIB CY compactification, hypermultiplets descend from two different sources. The
first one is the universal hypermultiplet, whose bosonic content is comprised of
four real scalar fields: the dilaton $\phi$ and the $C_0$-form, which combine
into the axio-dilaton $\tau=C_0+ie^{-\phi}$, and the two axions $b^0$ and $c^0$, dual
to the four-dimensional components of the NS-NS and R-R two-forms $B$ and
$C_2$, respectively. The remaining hypermultiplets are then parameterised by
$h^{1,1}$ different complexified K\"ahler moduli, defined as 
\begin{align}
	z^a = M_s^2 \int_{\gamma^a} B+iJ\,,
\end{align}
where $J$ stands for the CY K\"ahler form, $\text{Span}(\gamma^a)=H_2(X, \IZ)$ is a basis of integral 
two-cycles, and $M_s=(2\pi\sqrt{\alpha'})^{-1}$ the string scale. To define
\emph{bona fide} hypermultiplets these moduli partner with two axions,
$c^a$ and $d^a$, related to the R-R potentials $C_2$ and $C_4$ via
\begin{align}
   c^a = M_s^2 \int_{\gamma^a} C_2\, , \qquad \qquad  
   d_2^a = M_s^2\int_{\gamma^a} C_4\, ,
\end{align}
with $d^a$ being the four-dimensional dual of $d_2^a$. The vacuum expectation values of all these $4(h^{1,1}+1)$ real scalar fields parameterise the hypermultiplet moduli space $\mathcal{M}_\text{HM}$, that can be summarised as: 
%\begin{table}[h]
\begin{center}
\begin{tabular}{llll}
universal hypermultiplet & & $\tau = C_0 + i e^{-\phi}$, & $b^0$, \quad $c^0\, ,$\\
$h^{1,1}$ hypermultiplets & & $z^a  = b^a + it^a$, & $c^a$ ,\quad  $d^a\, .$
\end{tabular}
\end{center}
%\end{table}

In the following we choose to work in the so-called string frame, for which the ten-dimensional supergravity action takes the form
\begin{align}
	S_{10D} = 2\pi M_s^8\int_{\mathbb{R}^{1,3}\times X} d^{10}x \sqrt{-G} e^{-2\phi} R_{(10)} + \dots 
\end{align}
As we are interested in paths in the moduli space leading to classical infinite
distances, we focus on the non-periodic coordinates, the saxions:
\begin{align}
	\tau_2 = \text{Im}\;\tau= e^{-\phi} \,,\qquad t^a \,, \qquad a=1,\dots h^{1,1}(X)\,,
\end{align}
and for simplicity we will only consider trajectories for which the vacuum expectation values of all axions vanish.

At the classical level, the hypermultiplet moduli space metric reads \cite{Ferrara:1989ik}:
\begin{align}\label{metric}
	\mathrm{d}s^2_{{\cal M}_{\rm HM}}=\frac{1}{2} \left(\mathrm{d}\varphi_4\right)^2 + g_{a \bar b} \mathrm{d}z^a \mathrm{d}\bar z^b + \text{(axions)} \,,
\end{align}
where $\varphi_4$ stands for the 4d dilaton defined as
\begin{align}\label{phi4}
	e^{-2\varphi_4} = \frac{1}{2}\tau_2^2\, \mathcal{V}(t^a)\,,
\end{align}
with $\mathcal{V}$ the volume of $X$. This designates ${\varphi_4}$ as the four-dimensional physical field, rather than its ten-dimensional cousin $\tau_2$. It moreover defines a preferred
coordinate for the classical metric of the hypermultiplet moduli space, as it factorises from the K\"ahler moduli $z^a$. The classical metric $g_{a \bar b}$ for the latter can be obtained from a prepotential, and can be identified with that of the vector multiplet sector $\mathcal{M}_\text{VM}^{\rm IIA}$ in type IIA compactified on the same Calabi--Yau.

Having summarised the classical structure of $\mathcal{M}_\text{HM}$, we turn our attention to regions at its boundaries. More precisely, we consider paths of infinite distance in the saxionic directions $\varphi_4$ and $t^a$, aiming to understand the metric structure around its endpoints. Particularly interesting are those paths in which the classical 4d Planck mass 
\begin{align}\label{Planckmass}
	M_\text{Pl}^2
	= 4\pi e^{-2\varphi_4} M_s^2
	= 2\pi \tau_2^2 \,\mathcal{V}(t^a) M_s^2\,,
\end{align}
is kept fixed with respect to the string scale $M_s$. Indeed, such a constraint amounts to keep the 4d dilaton fixed, so that classically one moves in a copy of the vector multiplet $\mathcal{M}_\text{VM}^{\rm IIA}$ obtained from type IIA compactified on $X$. Moreover, the D-instanton actions take the form 
\begin{eqn}
S_{\bf{k}} = {2\pi}e^{-\varphi_4} e^{{\cal K}/2} | Z_{\bf{k}} | + 2\pi i \text{(axions)}\, ,
\label{Smk2}
\end{eqn}
where $e^{{\cal K}/2} | Z_{\bf{k}}|$ is a normalised central charge
function that only depends on the K\"ahler moduli $z^a$ and that, when we
compactify type IIA on $X$, corresponds to the mass of the D-particles in
4d Planck mass units, see e.g. \cite{Corvilain:2018lgw}.
Therefore, as stressed in \cite{Marchesano:2019ifh,Grimm:2019wtx}, one may use
the classification of infinite distance points performed in
\cite{Grimm:2018ohb,Grimm:2018cpv,Corvilain:2018lgw} to see that at endpoint
regions of infinite distance geodesics with $\varphi_4$ constant, towers of
instantons will develop an exponentially fast vanishing action. As such, they
will significantly modify the classical metric, to the point that they could
even remove the infinite distance at the quantum level
\cite{Marchesano:2019ifh}. 

Recently, an alternative classification of infinite distance geodesics in
$\mathcal{M}_\text{VM}^{\rm IIA}$ has been carried out in \cite{Lee:2019wij}. A
key aspect of this classification is to distinguish those infinite distance
paths along which the spectrum of four-dimensional particles signals a
decompactification limit, against those where this does not happen. The latter,
dubbed equi-dimensional limits, were conjectured to be dual to a
weakly-coupled string theory, described by either a type II or heterotic string
asymptotically tensionless along the path \cite{Lee:2019wij}.

One of the main purposes of this paper is to test the Emergent String
Conjecture of \cite{Lee:2019wij} beyond the vector multiplet moduli space of a
Calabi--Yau. In particular, this section and the next will confront it with
different infinite distance paths in the hypermultiplet moduli space
$\mathcal{M}_\text{HM}$ of type IIB on a Calabi--Yau $X$, and study which
insight this brings into the analysis performed in \cite{Marchesano:2019ifh}. 

From the discussion above, it is clear that a path in which the conjecture is
trivially satisfied is $\varphi_4 \rightarrow - \infty$. There, keeping all the
K\"ahler moduli constant, we are led to the weak-coupling region $\tau_2
\rightarrow \infty$ in which the perturbative, D-brane and NS5-brane instanton
corrections are irrelevant. Finally, from \eqref{Planckmass} it is trivial to
see that $\frac{M_s}{M_\text{Pl}} \raw 0$, and so the weakly-coupled string
realising the Emergent String Conjecture is nothing but the type IIB
fundamental string. The opposite limit can be seen as an S-dual analogue. 

As mentioned above a more interesting set of paths are those for which $\varphi_4$ is constant, and so quantum corrections induced by D-instantons will significantly modify the metric. Contrary to paths in vector multiplet moduli space, now the only obvious candidates to yield towers of massless particles are Kaluza--Klein modes and string winding modes. As a result one can build many more examples of equi-dimensional limits, providing further non-trivial tests to the Emergent String Conjecture. 

%%%%%%%%%%%%%%%%%%%%%%%%%%%%%%%%%%%%%%%%%%%%%%%%%
%%%%%%%%%%%%%%%%%%%%%%%%%%%%%%%%%%%%%%%%%%%%%%%%%
\subsubsection*{The D1-brane Limit}
%%%%%%%%%%%%%%%%%%%%%%%%%%%%%%%%%%%%%%%%%%%%%%%%%
%%%%%%%%%%%%%%%%%%%%%%%%%%%%%%%%%%%%%%%%%%%%%%%%%

One particular path that has been investigated in detail in
\cite{Marchesano:2019ifh} corresponds to the large-volume
$\mathcal{V}\rightarrow \infty$, strong-coupling, $\text{Im}\;\tau \rightarrow
0$, limit. This trajectory can be parameterised by a variable $\sigma\in
\mathbb{R}$, upon imposing the following scalings to the saxionic coordinates:
\begin{align}\label{D1limit}
	\text{D1: }\qquad 
	\tau_2(\sigma)=\tau_2(0) \, e^{-\frac{3}{2}\sigma} \,,\qquad
	t^a(\sigma)=t^a(0)\, e^{\sigma}\,,\qquad 
	\sigma\rightarrow \infty\,, 
\end{align}
where $t^a(0)$, $\tau_2(0)$ denote the starting point of the path in the
interior of the moduli space. At the classical level, these scalings
correspond to a constant 4d dilaton, thus ensuring that the
ratio between the Planck mass and the string scale does not change:
\begin{equation}
	\frac{M_s}{M_\text{Pl}}\sim \text{const}\,.
\end{equation}

While the Planck mass remains constant along the path, other mass scales might
of course vary. One of them is the Kaluza--Klein scale 
\begin{equation}
	\frac{M_\text{KK}^2}{M_\text{Pl}^2} \sim e^{-\sig}\, ,
\end{equation}
which one might think leads to a decompactification limit. However, this is not
so if one takes into account the tension of D1-strings extended along four
dimensions, which also vanishes asymptotically. In Planck mass units, we find:
\begin{align}\label{classicalD1Tension}
	\frac{T_\text{D1}}{M_\text{Pl}^2} = \tau_2\,\frac{M_s^2}{M_\text{Pl}^2}\sim e^{-\frac{3}{2}\sigma} \longrightarrow 0\,,  
\end{align}
and it therefore decreases faster than the Kaluza--Klein scale. Following the
reasoning of \cite{Lee:2019wij}, this limit qualifies as
equi-dimensional, and the obvious candidate for the emergent string is the
above D1-brane. Indeed, one can check that any other 4d string
obtained by dimensional reduction will be heavier. In the following, as
D1-strings are the lightest strings becoming tensionless in this limit, we will
dub the path \eqref{D1limit} the \textbf{D1 limit}.

While the presence of these tensionless strings were already noted in
\cite{Marchesano:2019ifh}, its significance for the low energy physics was not
investigated. The associated effective theory was then obtained by truncating
the spectrum of the fundamental type IIB string to its zero modes and
integrating out all massive states. The appearance of a tensionless string that
is not the fundamental string therefore signals a breakdown of the theory, as a
tower of states with masses below the cutoff of the original EFT, given by
$M_s$, arises. To capture the four-dimensional physics, one should therefore
look at the effective theory that can be obtained from the D1-string instead of
the effective theory of the fundamental string we started with. This
philosophy is in line with the idea of emergent strings at points of infinite
distance in moduli space \cite{Lee:2018urn,Lee:2019xtm,Lee:2019wij}.

To get a handle on the effective theory coming from the D1-string, we can make
use of S-duality, which in type IIB exchanges D1- and fundamental strings.
Thus the effective theory that governs the D1 limit \eqref{D1limit} is just a
variant of a compactification of the usual type IIB supergravity. The original
limit is however subject to a large number of non-perturbative instanton
corrections which significantly modify the original moduli space metric. As we
will see, a similar statement will hold after S-duality.  Hence, any classical
statements made in this section will have to be reconsidered when taking
quantum corrections into account.

%%%%%%%%%%%%%%%%%%%%%%%%%%%%%%%%%%%%%%%%%%%%%%%%%
%%%%%%%%%%%%%%%%%%%%%%%%%%%%%%%%%%%%%%%%%%%%%%%%%
\subsubsection*{S-duality and Fundamental Strings}
%%%%%%%%%%%%%%%%%%%%%%%%%%%%%%%%%%%%%%%%%%%%%%%%%
%%%%%%%%%%%%%%%%%%%%%%%%%%%%%%%%%%%%%%%%%%%%%%%%%

To understand the low energy physics in the D1 limit, one can use an
S-duality transformation to obtain a limit in which fundamental strings become
tensionless. Here, the four-dimensional EFT inherits S-duality from the
$SL(2,\mathbb{Z})$ symmetry of type IIB string theory, which amounts to
\begin{align}\label{Sduality}
	\tau \longrightarrow -\frac{1}{\tau}\,,\qquad
	t^a \longrightarrow |\tau| t^a \,,\qquad
	\begin{pmatrix}
		b^a\\c^a
	\end{pmatrix}
	\longrightarrow
	\begin{pmatrix}
		-c^a\\b^a
	\end{pmatrix}\,. 
\end{align}
Importantly, the 4d dilaton \eqref{phi4} is not invariant under
this transformation, $e^{-2\varphi_4}\to \tau_2^{-1}e^{-2\varphi_4}$, and so
the quotient $M_s/M_\text{Pl}$ will vary from one frame to the other.  To keep
track of the changes, let us denote the string scale after S-duality by
${M_s'}$ which---as reviewed in appendix \ref{app:PlanckMass}---is related to the
original string scale by ${M_s'}^2 =\tau_2 M_s^2$. Unsurprisingly, this
resembles the tension of D1-strings in the D1 limit as upon S-duality those are
exchanged with fundamental strings. Applying \eqref{Sduality} to the path
\eqref{D1limit}, one finds the following path in moduli space
\begin{align}\label{F1limit}
    \text{F1: }\qquad
    \tau'_2(\sigma)=\tau'_2(0) \, e^{+\frac{3}{2}\sigma} \,,\qquad 
    {t'}^a(\sigma)={t'}^a(0)\, e^{-\frac{\sigma}{2}}\,,\qquad 
     \sigma\rightarrow \infty\,, 
\end{align}
along which the ratio $M_s'/M_\text{Pl}$ no longer remains constant but instead
decreases. Indeed, in the transformed limit \eqref{F1limit} we thus get a
tensionless fundamental string
\begin{align}
	\frac{T'_\text{F1}}{M_\text{Pl}^2} = \frac{ {M_s'}^2}{M_\text{Pl}^2}= e^{2\varphi_4'} \sim e^{-\frac{3}{2}\sigma}\longrightarrow 0\,, 
\end{align}
The fundamental string becoming tensionless fastest, we refer to
\eqref{F1limit} as the \textbf{F1 limit}. Since this path in moduli space
drives us to a region of weak coupling and small volume, we again expect this
limit to be heavily quantum corrected. In particular, the small-volume point
might not belong to the quantum K\"ahler moduli space. We
come back to this issue and its implications in the next section. 

%%%%%%%%%%%%%%%%%%%%%%%%%%%%%%%%%%%%%%%%%%%%%%%%%
%%%%%%%%%%%%%%%%%%%%%%%%%%%%%%%%%%%%%%%%%%%%%%%%%
\subsubsection*{T-dualities and D3-branes}
%%%%%%%%%%%%%%%%%%%%%%%%%%%%%%%%%%%%%%%%%%%%%%%%%
%%%%%%%%%%%%%%%%%%%%%%%%%%%%%%%%%%%%%%%%%%%%%%%%%

It was shown in \cite{Lee:2019wij} that to reach an infinite distance point
while keeping a constant volume, the Calabi--Yau manifold must exhibit either
an elliptic, K3- or a $T^4$-fibration, such that the fibre shrinks to zero
size in that limit. Before delving into quantum corrections, we would like to
consider a limit of this sort, which moreover is related to the previous two
by some duality.

On the one hand, for K3- and $T^4$-fibrations a possible duality to the D1
limit holding at the quantum level is rather difficult to establish in general,
if existent at all. On the other hand, the case of an elliptic fibration has a
well-defined duality transform where the geometric limit of vanishing fibre can
be related to the D1 limit---and by extension to the F1 limit. As such, one
may use it as a starting point to investigate other (quantum-corrected)
infinite distance limits in hypermultiplet moduli space in the spirit of
\cite{Lee:2019wij}.

To establish such a duality, we will assume that the Calabi--Yau $X$ admits an
elliptic fibration over a two complex-dimensional K\"ahler base, $B_2$:
\begin{equation}
	\pi:~ X \longrightarrow B_2\,,
\end{equation}
with fibres denoted by $\mathcal{E}$. As we are interested in well-behaved
compactifications, we require the elliptic fibration to be smooth so that the
base admits at most divisors with singularities of type $I_1$ in the
Kodaira--N\'eron classification.

We can then divide the set of divisors generating the K\"ahler cone into vertical divisors
\begin{align}\label{verticaldivisor}
    D_\alpha = \pi^* \tilde{D}_\alpha\,, \qquad \alpha=2, \dots h^{1,1}(B_2)+1\,,
\end{align}
obtained by pulling back the K\"ahler cone generators $\tilde{D}_\alpha$ of
the base to the full fibration, and the divisor, 
\begin{align}\label{sectiondivisor}
    D_1 = S_0 + \pi^* c_1(B_2) \,,
\end{align}
related to the zero section $S_0$ and the first Chern class of the base. The Poincar\'e dual of these divisors, $J_\alpha$ and $J_1$
respectively, are used to expand the K\"ahler form of the elliptic fibration in terms
of $h^{1,1}(B_2)+1$ moduli:
\begin{align}
	J=t^1 J_1 + \sum_{\alpha=2}^{h^{1,1}(B_2)+1} t^\alpha J_\alpha \,, 
\end{align}
where the first modulus $t^1$ parametrises the volume of the fibre.

The limit we want to consider is then the regime where the volume of the
elliptic fibre vanishes while keeping the ratio between the Planck mass and the
string scale fixed.  In the original D1 limit \eqref{D1limit} all moduli are
blown up, and we can therefore obtain a shrinking fibre by performing a double
T-duality:
\begin{align}
    t^1\rightarrow \frac{1}{t^1}\,.\label{doubleTduality}
\end{align}
However, for non-trivial fibrations we have to account for possible
$I_1$-singularities and curvature of the base. This simple double T-duality
should then be replaced by a certain Fourier--Mukai transform\footnote{Such a
	transform has already been used in the SDC context in
\cite{Corvilain:2018lgw} to relate asymptotically massless D-brane particle
states at large fibre volume to the corresponding states at small volume.}
\cite{Andreas:2004uf}, which has been shown to corresponds to a monodromy
action on the K\"ahler moduli space \cite{Cota:2019cjx}.

In terms of the complexified K\"ahler moduli, the Fourier--Mukai transform acts
as \cite{Schimannek:2019ijf}:
\begin{align}\label{FMTransform}
	z^1\longrightarrow -\frac{1}{z^1}\,,\qquad 
	%\tau_2\longrightarrow |z^1|\tau_2\,,\qquad 
	z^\alpha\longrightarrow z^\alpha + \frac{1}{2}K^\alpha z^1+\frac{K^\alpha}{2z^1} \,, 
\end{align}
where $K^\alpha$ denotes the coefficients of the first Chern class of the base
expanded in the K\"ahler cone basis, $c_1(B_2) = K^\alpha J_\alpha$, and
account for the non-trivial fibration structure. To keep a constant
string-to-Planck-scale ratio the dilaton further needs to be adjusted. As shown
in appendix \ref{app:PlanckMass} the correct transformation is given to leading
order by:
\begin{equation}\label{transformtau2}
	\tau_2\longrightarrow |z^1|\tau_2 +\dots
\end{equation}
The ellipses correspond to terms involving $K^\alpha$ which will not play any
r\^ole in the limit we consider. In this new frame the dual path indeed
corresponds to shrinking the elliptic fibre and blowing up the volume of the
base while going to strong coupling, as can be seen from the new scalings:
\begin{align}\label{D3limit}
	\text{D3: }\qquad
	\tau_2''= e^{-\frac{1}{2}\sigma}\tau_2''(0)\,,\qquad
	{t''}^1(\sigma) = e^{-\sigma} {t''}^1(0) \,,\qquad 
	{t''}^\alpha= e^{\sigma} {t''}^\alpha(0)\,.
\end{align}
Again, to facilitate the comparison quantities in this frame will be double primed. 
Note that the limit we obtain corresponds to a so-called limit with co-scaling in the nomenclature of 
\cite{Lee:2019wij}: while the string-to-Planck-scale ratio remains constant, the
volume blows up and is compensated by a scaling to strong coupling. 

Moreover, similarly to what happened with S-duality the lightest string is not
a D1-brane in that limit. Under the double T-duality, D1-branes are mapped to
D3-branes wrapping the elliptic fibre, and these are the lightest ones becoming
asymptotically tensionless. Indeed, in the limit \eqref{D3limit} one obtains:
\begin{align}\label{tensionD3class}
	\frac{T''_{\text{D3}|_\mathcal{E}}}{M_\text{Pl}^2}
	= \tau_2'' {t''}^1 \left(\frac{M_s''}{M_\text{Pl}}\right)^2 
	\sim e^{-\frac{3}{2}\sigma} \longrightarrow 0\,.
\end{align}
We use the notation $\left.T''_\text{D3}\right|_{\mathcal{E}}$ to emphasise
that in the dual frame, tensionless strings at the endpoint come from D3-branes
wrapping the fibre. As any other string can be shown to remain heavier, we again
use the nomenclature denoting a frame by its asymptotically lightest
string and refer to \eqref{D3limit} as the {\bf D3 limit}. 

\begin{table}[htb]
    \centering
    \begin{tabular}{|c|c|c|c|}\hline 
        Limit& $\tau_2$ & $t^1$ & $t^\alpha$ \\ \hline \hline 
        D1 & $e^{-3/2\sigma}$ & $e^\sigma$ & $e^\sigma$ \\ \hline 
        F1 & $e^{3/2\sigma}$ & $e^{-\sigma/2}$ & $e^{-\sigma/2}$ \\ \hline  
        D3 & $e^{-1/2\sigma}$ & $e^{-\sigma}$ & $e^\sigma$ \\ \hline 
    \end{tabular}
    \caption{Scaling of the saxionic fields in the different duality frames.}
    \label{tab::Einsteinscalings}
\end{table} 

\begin{figure}[t]
	\centering
	\resizebox{.5\textwidth}{!}{\Huge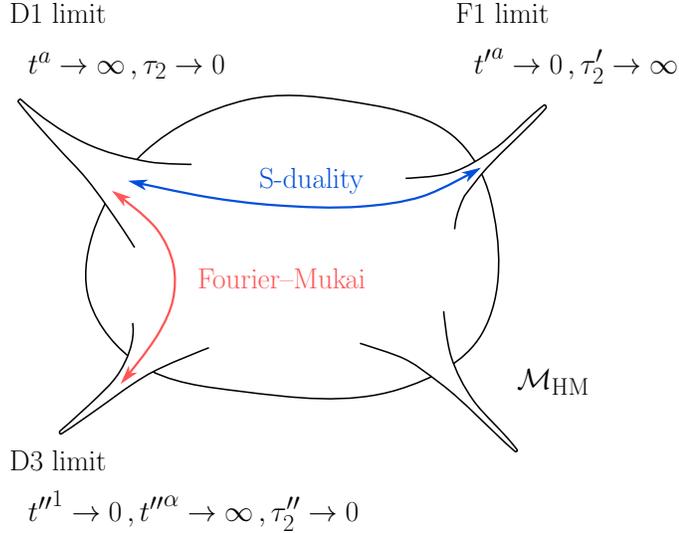}
	\caption{Different classical infinite distance limits discussed in section \ref{sec:setup}, taken along the hypermultiplet sector of type IIB Calabi--Yau compactifications. The name of the frame refers to the lightest string in each of the limits.}
	\label{fig:moduliSpace}
\end{figure}

To summarise, we have related in this section three different classical limits
where different types of 4d strings become tensionless by using
part of the type IIB U-duality group. The scalings of the moduli in all three
regimes are summarised in table \ref{tab::Einsteinscalings}, and the dualities
relating them illustrated in figure \ref{fig:moduliSpace}. As discussed in the
next section, the classical approximation breaks down along the path in each
frame, and one has to take quantum corrections into account.
As we will see, the dominant quantum effects are different in each frame and
they will play a key r\^ole in understanding the four-dimensional EFT physics.
In particular they will allow us to refine previous arguments in
\cite{Marchesano:2019ifh} and to show how the emergent strings proposal of
\cite{Lee:2019wij} is realised in our setup. 

%%%%%%%%%%%%%%%%%%%%%%%%%%%%%%%%%%%%%%%%%%%%%%%%%
%%%%%%%%%%%%%%%%%%%%%%%%%%%%%%%%%%%%%%%%%%%%%%%%%
%%%%%%%%%%%%%%%%%%%%%%%%%%%%%%%%%%%%%%%%%%%%%%%%%
\section{Quantum Corrections to Moduli Space Geometry}\label{sec:QuantumCorrections}
%%%%%%%%%%%%%%%%%%%%%%%%%%%%%%%%%%%%%%%%%%%%%%%%%
%%%%%%%%%%%%%%%%%%%%%%%%%%%%%%%%%%%%%%%%%%%%%%%%%
%%%%%%%%%%%%%%%%%%%%%%%%%%%%%%%%%%%%%%%%%%%%%%%%%

Having established the classical infinite distance limits that we want to
investigate, we now address how quantum corrections affect the hypermultiplet
moduli space along them. As we will see, in each of the limits a different set
of quantum corrections will significantly change the classical moduli space
geometry, along the lines of \cite{Marchesano:2019ifh}. As we are particularly
interested in testing the Emergent String Conjecture in this setup, parallel to
this analysis we will keep track of how the different 4d string tensions behave
along the limits. In this sense, a useful organising principle to compare
non-perturbative quantum corrections with string tensions will be the
following: whenever a non-perturbative correction becomes significant, the
tension of a certain 4d string becomes lighter than $M_s^2$, and the other way
around \footnote{This reflects into different regions of the moduli space as shown in Fig. \ref{fig:regions} below.}. This will allow us to sharpen the arguments of
\cite{Marchesano:2019ifh}, and to see how each of the different limits---which remain of infinite distance---leads to a
weakly-coupled type II string theory emerging at the trajectory's endpoint. Notice that as a consequence of this emergence, at the far end of
the trajectory the hypermultiplet metric will have to be of the classical form
\eqref{metric}. This simplification will also be crucial in
interpreting our results.

%%%%%%%%%%%%%%%%%%%%%%%%%%%%%%%%%%%%%%%%%%%%%%%%%
%%%%%%%%%%%%%%%%%%%%%%%%%%%%%%%%%%%%%%%%%%%%%%%%%
\subsection{Quantum Corrections in the D1 Limit}\label{ss:D1limit}
%%%%%%%%%%%%%%%%%%%%%%%%%%%%%%%%%%%%%%%%%%%%%%%%%
%%%%%%%%%%%%%%%%%%%%%%%%%%%%%%%%%%%%%%%%%%%%%%%%%

Quantum corrections along the D1 limit \eqref{D1limit} were considered in
detail in \cite{Marchesano:2019ifh}. Particular attention was given to the
corrections to the  hypermultiplet metric, which in general can be determined
by studying the quaternionic-K\"ahler structure of $\mathcal{M}_\text{HM}$ in
twistor space (see the reviews \cite{Alexandrov:2011va,Alexandrov:2013yva} and references therein). In the
case of the D1 limit this structure simplifies, because corrections due to D3-,
D5-and NS5-brane instantons can be safely neglected, and one can encode the
corrections to the metric in a contact potential \cite{deWit:2006gn} first
computed in \cite{RoblesLlana:2006is}. The structure of this contact potential
$\chi$ reads 
\begin{equation}
\chi = \chi_{\rm cl} + \chi_{\rm corr}\, ,
\label{contact}
\end{equation}
where
\begin{equation}
\chi_{\rm cl} = \frac{1}{12} \tau_2^2  {\cal K}_{abc}t^at^bt^c 
\label{chiclas}
\end{equation}
is the classical contribution to the potential, with $\mathcal{K}_{abc}$ the
triple-intersection numbers of $X$ and
\begin{equation}
\chi_{\rm corr} = \frac{\tau_2^2}{8(2\pi)^3} \sum_{\bf{k}\geq 0} n_{\bm{k}}^{(0)} \sum_{(m,n)\in \mathbb{Z}^2\backslash 0}  \frac{1+ 2\pi |m\tau + n|k_at^a}{|m\tau + n|^3}\, e^{-S_{m,n}^{\bm{k}} }\, ,
\label{chicorr}
\end{equation}
the relevant perturbative and non-perturbative corrections. Here $\bm{k}$ is a
vector of $h^{1,1}$ entries $k_a \in \IN$, such that $k_a \gamma^a$ spans
homology classes in $H_2^+(X,\mathbb{Z})$. Terms with $\bm{k}\neq \bm{0}$
represent corrections coming from Euclidean $(m,n)$-strings wrapping
two-cycle classes with non-vanishing genus-zero Gopakumar--Vafa invariant
$n_{\bm{k}}^{(0)}$, with action
\begin{equation}\label{Smn}
	S_{m,n}^{\bm{k}} = 2\pi k_a \left(|m\tau + n| t^a  - imc^a - inb^a \right)\, .
\end{equation}
The contribution from the term $\bm{k} = \bm{0}$ corresponds to the D(-1)-brane
corrections, together with the one-loop corrections on $g_s$ and $\alpha'$, if
one sets $n_{\bm{k}=\bm{0}}^{(0)}= - \chi_E(X)$, that is to (minus) the Euler
characteristic of $X$. 

It was found in \cite{Marchesano:2019ifh} that:

\begin{itemize}

\item[{\it i)}] $\chi \sim \chi_{\rm corr} \gg \chi_{\rm cl}$ as one proceeds
along the D1 limit. Since $\chi_{\rm cl}$ is nothing but the classical 4d
dilaton \eqref{phi4}, one can see $\chi$ as its quantum-corrected version \cite{Alexandrov:2008gh}, and
in particular the quantum-corrected string-to-Planck-scale ratio. Therefore,
even if the D1 limit was designed to keep such a ratio constant at the
classical level, D-instanton corrections make it blow up. One can moreover
identify this growth with the number of instanton species $N_{sp}(\sigma)$ that acquire a small action at a given point $\sigma$ along the trajectory
\eqref{D1limit}, so that
\begin{equation}\label{Msp}
	M_s \sim \frac{M_\text{Pl}}{\sqrt{N_{sp}(\sigma)}}\, ,
\end{equation}
and the string scale can be interpreted as some sort of species scale for D-instantons. 

\item[{\it ii)}] Even if in general the presence of $\chi_{\rm corr}$ distorts
the hypermultiplet metric away from its classical form \eqref{metric}, as one
proceeds along the D1 limit a simplification occurs, in the sense that the
scalings of the different metric factors coincide with those of
\begin{align}\label{metricD1}
	\mathrm{d}s^2_{{\cal M}_{\rm HM}}=\frac{1}{2} \left(\mathrm{d}{\rm log} \chi \right)^2 + g_{a \bar b} \mathrm{d}z^a \mathrm{d}\bar z^b + \text{(axions)} \,,
\end{align}
\begin{equation}
g_{a\bar{b}} = \partial_{z^a}\partial_{\bar{z}^{b}} K \quad \quad {\rm with} \quad \quad K = - {\rm log}\, \chi \, ,
\label{qcmetric}
\end{equation}
where $\chi$ is to be evaluated for large values of $\sigma$ in \eqref{D1limit}.

\end{itemize}

To determine such scalings one needs to evaluate the dependence of $\chi$ and its derivatives with the trajectory coordinate $\sigma$. The strategy followed in \cite{Marchesano:2019ifh} made use of the standard split of $\chi_{\rm corr}$  into three distinct sources \cite{RoblesLlana:2007ae},
\begin{equation}\label{chisplit}
	\chi_\text{corr} = \chi_\text{pert} + \chi_\text{WS} + \chi_\text{D}\,.
\end{equation}
Here the first term is the perturbative piece, interpreted as the one-loop corrections in $\alpha^\prime$ and $g_s$,
\begin{equation}\label{chiPert}
	\chi_\text{pert} = -\frac{\chi_E(X)}{4(2\pi)^3}\left(\zeta(3)\tau_2^2+\frac{\pi^2}{3}\right)\,,
\end{equation}
and controlled by the Euler density of the compactification manifold $\chi_E(X)$. The second contribution encodes worldsheet instantons associated to any effective two-cycle, and can be written as a sum weighted by genus-zero Gopakumar--Vafa invariants  $n_{\bm{k}}^{(0)}$:
\begin{equation}\label{chiWS}
	\chi_\text{WS} = \frac{\tau^2_2}{4(2\pi)^3}\sum_{\bm{k}>\bm{0}} n_{\bm{k}}^{(0)}\text{Re}\left(\text{Li}_3(e^{2\pi i k_az^a})+2\pi k_at^a\text{Li}_2(e^{2\pi i k_az^a})\right)\,,
\end{equation}
with $\text{Li}_s(x)=\sum_{r>0}r^{-s}x^r$ the polylogarithm function. Finally, the last term corresponds to D1- and D(-1)-instanton corrections:
\begin{gather}\label{chiD}
	\chi_\text{D}  = \frac{\tau_2}{8\pi^2}\sum_{k_\Lambda\neq0} n_{\bm{k}}^{(0)}\sum_{m>0}\frac{|k_\Lambda z^{\Lambda}|}{m}\cos(2\pi m k_\Lambda \zeta^\Lambda) K_1(2\pi m |k_\Lambda z^\Lambda|\tau_2)\,,\\
	z^\Lambda=(1,z^a)\,,\qquad \zeta^\Lambda=(\tau_1,\tau_1b^a-c^a)\, ,\nonumber
\end{gather}
with $K_1$ the modified Bessel function. The sum is performed over  $k_\Lambda = (k_0,k_a)\neq0$, such that $k_0 \in \IZ$ and $k_a\in \IN$,  interpreted as the D(-1) and D1-instanton charges, respectively.

As one proceeds along the trajectory \eqref{D1limit} the three terms in \eqref{chisplit} will have very different behaviours. This is already to be expected from the scaling of the different instanton actions involved, namely
\begin{subequations}
\begin{align}
	S^{\rm D(-1)}_{k_0}=\tau_2 k_0 & \stackrel{~\eqref{D1limit}~}{\longrightarrow} e^{-\frac{3}{2}\sigma}\,, \label{D1D1scaling} \\
	S^{\rm D1}_{\bm{k}}=\tau_2 k_a z^a& \stackrel{~\eqref{D1limit}~}{\longrightarrow} e^{-\frac{1}{2}\sigma} \,,\\
	S^{\rm WS}_{\bm{k}}=k_a z^a & \stackrel{~\eqref{D1limit}~}{\longrightarrow} e^{+\sigma} \,,
\end{align}
\end{subequations}
and so the corrections to the moduli space metric mainly arise due to D1- and
D(-1)-instantons. Indeed,
one can check that along the D1 limit the contributions of $\chi_\text{pert}$
and $\chi_\text{WS}$ are negligible compared to that of $\chi_\text{D}$, and
that the growth of the latter can be described by using that for small
arguments, the modified Bessel function behaves as 
\begin{align}
    K_1(x)\sim\frac{1}{x}\qquad \text{as}\; x\rightarrow 0\,.
\end{align}
At the end of the day, one obtains that the significant contributions to the
contact potential take the form
\begin{align}\label{limitchiD1}
	\chi \sim \frac{1}{12} \tau_2^2  {\cal K}_{abc}t^at^bt^c + \frac{1}{96\pi} \left[-\chi_E(X) + \sum_{{\bf k} > 0}^\Omega n_{\bf{k}}^{(0)}\right]\sum_{k_0}^{k_0^\text{max}} 1\,,
\end{align}
where the sum over $k_\Lambda = (k_0,k_a)$ is cut off by $k_0^\text{max}$ and
$\Omega$, which contain those instanton charges for which the condition
\begin{equation}\label{instantonCondition}
	\text{Re}(S^\text{D}_{k^\Lambda}) = 2\pi |k_0e^{-\frac{3}{2}\sigma} +k_ae^{-\frac{1}{2}\sigma}| \ll 1
\end{equation}
is satisfied. This cutoff is introduced to parametrize the approximation of the Bessel function in \eqref{limitchiD1} and can physically be interpreted as distinguishing instantons that contribute significantly to $\chi$ from those that are exponentially suppressed.  As mentioned previously, as we move along the trajectory
\eqref{D1limit}, a growing number of instantons acquire a small action and
hence the second term in \eqref{limitchiD1} eventually dominates over the
classical part. Finally, in \cite{Marchesano:2019ifh} the condition
\eqref{instantonCondition} was approximated as 
\begin{align}\label{approxeps}
	\text{Re}(S^\text{D}_{k^\Lambda}) = |k_0e^{-\frac{3}{2}\sigma} +k_ae^{-\frac{1}{2}\sigma}| \le e^{-\epsilon \sigma}\,,
\end{align}
for $\eps \in \IR$ some fixed parameter. This imposes the following cutoff on
the instanton sum
\begin{align}\label{chargeGrowthD1New}
     k_0\le e^{(3/2-\epsilon)\sigma} \,,\qquad
     k_a\le e^{(1/2-\epsilon)\sigma}\, .
\end{align}
Then,  whenever $\eps >0$ the hard cutoff on the action of the instantons in
the sum decreases gradually. This parallels the behaviour of wrapped D-brane
particles contributing to loop corrections of the moduli space metric of type
II vector multiplet moduli spaces, as considered in the emergence proposal
\cite{Grimm:2018ohb}. In that case, the gradually decreasing cutoff is
identified with the species scale, leading to a break down of the associated
effective theory. It was then found in \cite{Marchesano:2019ifh} that for $\eps
>0$ the quantum corrections to the metric remove the infinite distance, while
the path \eqref{D1limit} remains of infinite length for $\eps =0$.

In the following we would like to revisit the approximation
\eqref{instantonCondition}, in light of the connection between instantons of
low action and the 4d strings of low tension, and the r\^ole 
the latter play in the Emergent String Conjecture.

\subsubsection*{Instantons and Light Strings}

The two features found in \cite{Marchesano:2019ifh} related to \eqref{Msp} and
\eqref{metricD1} are very suggestive from the viewpoint of the Emergent String
Conjecture. However, the fundamental string associated to them is not the lightest 4d string,
and so it cannot realise the Emergent String Conjecture. Indeed, as already mentioned, along the
trajectory \eqref{D1limit} a second effect occurs, namely that the D1-brane
oscillation modes become lighter than those of the fundamental string. This
sort of effect is not a coincidence, and goes hand in hand with instanton
effects significantly modifying the classical metric. Indeed, the D1-brane
tension goes below the fundamental string one in the region $\tau_2 < 1$, which
is when D(-1)-brane instantons start to significantly modify the classical
metric, as captured by \eqref{instantonCondition}. This is both true
classically and after the corrections that change the string-to-Planck-scale
ratio as in \eqref{Msp} have been taken into account.

Similarly, a D3-brane wrapped on an internal two-cycle of $X$ will be seen as a
4d string whose tension in units of $M_s^2$ decreases as $t^a \tau_2 \sim
e^{-\sig/2}$, just like the action of a D1-brane instanton wrapped on the same
two-cycle. One can then separate the trajectory \eqref{metricD1} into three
regions according to the relevance of D-brane instantons and their
corresponding 4d strings:
%
%\begin{table}[h]
\begin{center}
\begin{tabular}{ccccc}
{\bf Region I} & \quad \quad & $T_{\rm F1} \, <\, T_{\rm D1}  \, <\, T_{\rm D3}$  & \quad \quad& $\chi_{\rm cl} \, > \chi_{\rm D}$ \\
{\bf Region II} & \quad \quad& $T_{\rm D1} \, <\, T_{\rm F1}  \, <\, T_{\rm D3}$ & \quad \quad& $\chi_{\rm D(-1)} \, > \chi_{\rm cl} \, > \chi_{\rm D1}$\\
{\bf Region III} & \quad \quad & $T_{\rm D1} \, <\, T_{\rm D3}  \, <\, T_{\rm F1}$ & \quad \quad& $\chi_{\rm D(-1)}, \chi_{\rm D1} \, > \chi_{\rm cl}$
\end{tabular}
\end{center}
%\end{table}

\begin{figure}[ht]
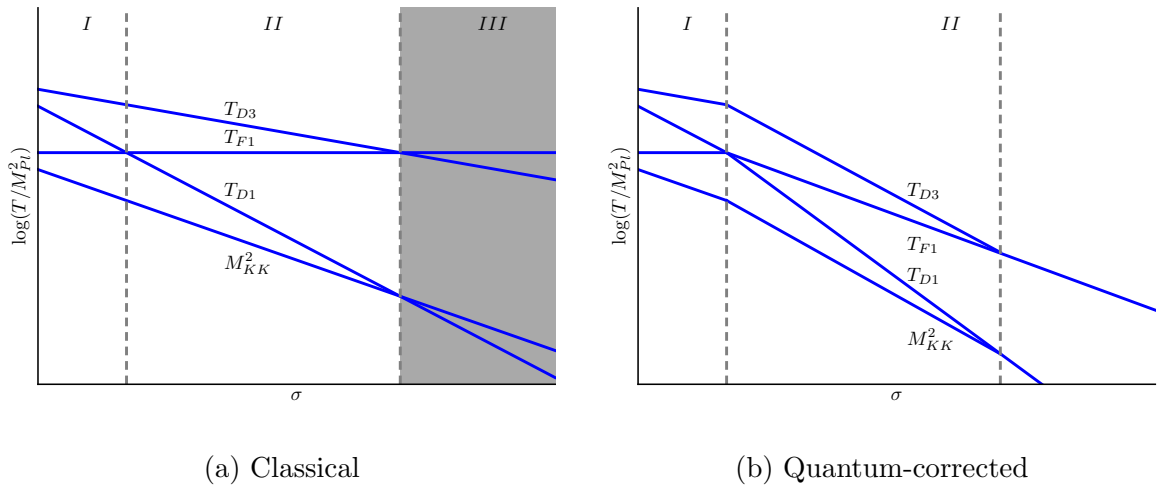

	\centering
	\begin{subfigure}[t]{0.49\linewidth}
		\centering
		\resizebox{\linewidth}{!}{
	        	%% Creator: Matplotlib, PGF backend
%%
%% To include the figure in your LaTeX document, write
%%   \input{<filename>.pgf}
%%
%% Make sure the required packages are loaded in your preamble
%%   \usepackage{pgf}
%%
%% Figures using additional raster images can only be included by \input if
%% they are in the same directory as the main LaTeX file. For loading figures
%% from other directories you can use the `import` package
%%   \usepackage{import}
%% and then include the figures with
%%   \import{<path to file>}{<filename>.pgf}
%%
%% Matplotlib used the following preamble
%%
\begingroup%
\makeatletter%
\begin{pgfpicture}%
\pgfpathrectangle{\pgfpointorigin}{\pgfqpoint{4.110671in}{3.074012in}}%
\pgfusepath{use as bounding box, clip}%
\begin{pgfscope}%
\pgfsetbuttcap%
\pgfsetmiterjoin%
\definecolor{currentfill}{rgb}{1.000000,1.000000,1.000000}%
\pgfsetfillcolor{currentfill}%
\pgfsetlinewidth{0.000000pt}%
\definecolor{currentstroke}{rgb}{1.000000,1.000000,1.000000}%
\pgfsetstrokecolor{currentstroke}%
\pgfsetdash{}{0pt}%
\pgfpathmoveto{\pgfqpoint{0.000000in}{0.000000in}}%
\pgfpathlineto{\pgfqpoint{4.110671in}{0.000000in}}%
\pgfpathlineto{\pgfqpoint{4.110671in}{3.074012in}}%
\pgfpathlineto{\pgfqpoint{0.000000in}{3.074012in}}%
\pgfpathclose%
\pgfusepath{fill}%
\end{pgfscope}%
\begin{pgfscope}%
\pgfsetbuttcap%
\pgfsetmiterjoin%
\definecolor{currentfill}{rgb}{1.000000,1.000000,1.000000}%
\pgfsetfillcolor{currentfill}%
\pgfsetlinewidth{0.000000pt}%
\definecolor{currentstroke}{rgb}{0.000000,0.000000,0.000000}%
\pgfsetstrokecolor{currentstroke}%
\pgfsetstrokeopacity{0.000000}%
\pgfsetdash{}{0pt}%
\pgfpathmoveto{\pgfqpoint{0.310278in}{0.279012in}}%
\pgfpathlineto{\pgfqpoint{4.010671in}{0.279012in}}%
\pgfpathlineto{\pgfqpoint{4.010671in}{2.974012in}}%
\pgfpathlineto{\pgfqpoint{0.310278in}{2.974012in}}%
\pgfpathclose%
\pgfusepath{fill}%
\end{pgfscope}%
\begin{pgfscope}%
\pgfpathrectangle{\pgfqpoint{0.310278in}{0.279012in}}{\pgfqpoint{3.700393in}{2.695000in}}%
\pgfusepath{clip}%
\pgfsetbuttcap%
\pgfsetroundjoin%
\definecolor{currentfill}{rgb}{0.662745,0.662745,0.662745}%
\pgfsetfillcolor{currentfill}%
\pgfsetlinewidth{0.000000pt}%
\definecolor{currentstroke}{rgb}{0.000000,0.000000,0.000000}%
\pgfsetstrokecolor{currentstroke}%
\pgfsetdash{}{0pt}%
\pgfpathmoveto{\pgfqpoint{2.901944in}{7.120166in}}%
\pgfpathlineto{\pgfqpoint{2.901944in}{-3.245219in}}%
\pgfpathlineto{\pgfqpoint{5.493610in}{-3.245219in}}%
\pgfpathlineto{\pgfqpoint{5.493610in}{7.120166in}}%
\pgfpathlineto{\pgfqpoint{5.493610in}{7.120166in}}%
\pgfpathlineto{\pgfqpoint{2.901944in}{7.120166in}}%
\pgfpathclose%
\pgfusepath{fill}%
\end{pgfscope}%
\begin{pgfscope}%
\definecolor{textcolor}{rgb}{0.000000,0.000000,0.000000}%
\pgfsetstrokecolor{textcolor}%
\pgfsetfillcolor{textcolor}%
\pgftext[x=2.160474in,y=0.223457in,,top]{\color{textcolor}\rmfamily\fontsize{10.000000}{12.000000}\selectfont \(\displaystyle \sigma\)}%
\end{pgfscope}%
\begin{pgfscope}%
\definecolor{textcolor}{rgb}{0.000000,0.000000,0.000000}%
\pgfsetstrokecolor{textcolor}%
\pgfsetfillcolor{textcolor}%
\pgftext[x=0.254723in,y=1.626512in,,bottom,rotate=90.000000]{\color{textcolor}\rmfamily\fontsize{10.000000}{12.000000}\selectfont \(\displaystyle \log(T/M_{Pl}^2)\)}%
\end{pgfscope}%
\begin{pgfscope}%
\pgfpathrectangle{\pgfqpoint{0.310278in}{0.279012in}}{\pgfqpoint{3.700393in}{2.695000in}}%
\pgfusepath{clip}%
\pgfsetrectcap%
\pgfsetroundjoin%
\pgfsetlinewidth{1.505625pt}%
\definecolor{currentstroke}{rgb}{0.000000,0.000000,1.000000}%
\pgfsetstrokecolor{currentstroke}%
\pgfsetdash{}{0pt}%
\pgfpathmoveto{\pgfqpoint{0.310278in}{2.270127in}}%
\pgfpathlineto{\pgfqpoint{4.009191in}{0.327395in}}%
\pgfpathlineto{\pgfqpoint{4.009191in}{0.327395in}}%
\pgfusepath{stroke}%
\end{pgfscope}%
\begin{pgfscope}%
\pgfpathrectangle{\pgfqpoint{0.310278in}{0.279012in}}{\pgfqpoint{3.700393in}{2.695000in}}%
\pgfusepath{clip}%
\pgfsetrectcap%
\pgfsetroundjoin%
\pgfsetlinewidth{1.505625pt}%
\definecolor{currentstroke}{rgb}{0.000000,0.000000,1.000000}%
\pgfsetstrokecolor{currentstroke}%
\pgfsetdash{}{0pt}%
\pgfpathmoveto{\pgfqpoint{0.310278in}{2.391202in}}%
\pgfpathlineto{\pgfqpoint{4.009191in}{1.743625in}}%
\pgfpathlineto{\pgfqpoint{4.009191in}{1.743625in}}%
\pgfusepath{stroke}%
\end{pgfscope}%
\begin{pgfscope}%
\pgfpathrectangle{\pgfqpoint{0.310278in}{0.279012in}}{\pgfqpoint{3.700393in}{2.695000in}}%
\pgfusepath{clip}%
\pgfsetrectcap%
\pgfsetroundjoin%
\pgfsetlinewidth{1.505625pt}%
\definecolor{currentstroke}{rgb}{0.000000,0.000000,1.000000}%
\pgfsetstrokecolor{currentstroke}%
\pgfsetdash{}{0pt}%
\pgfpathmoveto{\pgfqpoint{0.310278in}{1.816398in}}%
\pgfpathlineto{\pgfqpoint{4.009191in}{0.521244in}}%
\pgfpathlineto{\pgfqpoint{4.009191in}{0.521244in}}%
\pgfusepath{stroke}%
\end{pgfscope}%
\begin{pgfscope}%
\pgfpathrectangle{\pgfqpoint{0.310278in}{0.279012in}}{\pgfqpoint{3.700393in}{2.695000in}}%
\pgfusepath{clip}%
\pgfsetrectcap%
\pgfsetroundjoin%
\pgfsetlinewidth{1.505625pt}%
\definecolor{currentstroke}{rgb}{0.000000,0.000000,1.000000}%
\pgfsetstrokecolor{currentstroke}%
\pgfsetdash{}{0pt}%
\pgfpathmoveto{\pgfqpoint{0.310278in}{1.937474in}}%
\pgfpathlineto{\pgfqpoint{4.009191in}{1.937474in}}%
\pgfpathlineto{\pgfqpoint{4.009191in}{1.937474in}}%
\pgfusepath{stroke}%
\end{pgfscope}%
\begin{pgfscope}%
\pgfpathrectangle{\pgfqpoint{0.310278in}{0.279012in}}{\pgfqpoint{3.700393in}{2.695000in}}%
\pgfusepath{clip}%
\pgfsetbuttcap%
\pgfsetroundjoin%
\pgfsetlinewidth{1.505625pt}%
\definecolor{currentstroke}{rgb}{0.501961,0.501961,0.501961}%
\pgfsetstrokecolor{currentstroke}%
\pgfsetdash{{4.500000pt}{3.000000pt}}{0.000000pt}%
\pgfpathmoveto{\pgfqpoint{0.943642in}{0.279012in}}%
\pgfpathlineto{\pgfqpoint{0.943642in}{2.974012in}}%
\pgfusepath{stroke}%
\end{pgfscope}%
\begin{pgfscope}%
\pgfpathrectangle{\pgfqpoint{0.310278in}{0.279012in}}{\pgfqpoint{3.700393in}{2.695000in}}%
\pgfusepath{clip}%
\pgfsetbuttcap%
\pgfsetroundjoin%
\pgfsetlinewidth{1.505625pt}%
\definecolor{currentstroke}{rgb}{0.501961,0.501961,0.501961}%
\pgfsetstrokecolor{currentstroke}%
\pgfsetdash{{4.500000pt}{3.000000pt}}{0.000000pt}%
\pgfpathmoveto{\pgfqpoint{2.901944in}{0.279012in}}%
\pgfpathlineto{\pgfqpoint{2.901944in}{2.974012in}}%
\pgfusepath{stroke}%
\end{pgfscope}%
\begin{pgfscope}%
\pgfsetrectcap%
\pgfsetmiterjoin%
\pgfsetlinewidth{0.803000pt}%
\definecolor{currentstroke}{rgb}{0.000000,0.000000,0.000000}%
\pgfsetstrokecolor{currentstroke}%
\pgfsetdash{}{0pt}%
\pgfpathmoveto{\pgfqpoint{0.310278in}{0.279012in}}%
\pgfpathlineto{\pgfqpoint{0.310278in}{2.974012in}}%
\pgfusepath{stroke}%
\end{pgfscope}%
\begin{pgfscope}%
\pgfsetrectcap%
\pgfsetmiterjoin%
\pgfsetlinewidth{0.803000pt}%
\definecolor{currentstroke}{rgb}{0.000000,0.000000,0.000000}%
\pgfsetstrokecolor{currentstroke}%
\pgfsetdash{}{0pt}%
\pgfpathmoveto{\pgfqpoint{0.310278in}{0.279012in}}%
\pgfpathlineto{\pgfqpoint{4.010671in}{0.279012in}}%
\pgfusepath{stroke}%
\end{pgfscope}%
\begin{pgfscope}%
\definecolor{textcolor}{rgb}{0.000000,0.000000,0.000000}%
\pgfsetstrokecolor{textcolor}%
\pgfsetfillcolor{textcolor}%
\pgftext[x=1.642419in,y=2.005382in,left,base]{\color{textcolor}\rmfamily\fontsize{10.000000}{12.000000}\selectfont \(\displaystyle T_{F1}\)}%
\end{pgfscope}%
\begin{pgfscope}%
\definecolor{textcolor}{rgb}{0.000000,0.000000,0.000000}%
\pgfsetstrokecolor{textcolor}%
\pgfsetfillcolor{textcolor}%
\pgftext[x=1.642419in,y=1.627633in,left,base]{\color{textcolor}\rmfamily\fontsize{10.000000}{12.000000}\selectfont \(\displaystyle T_{D1}\)}%
\end{pgfscope}%
\begin{pgfscope}%
\definecolor{textcolor}{rgb}{0.000000,0.000000,0.000000}%
\pgfsetstrokecolor{textcolor}%
\pgfsetfillcolor{textcolor}%
\pgftext[x=1.642419in,y=2.206349in,left,base]{\color{textcolor}\rmfamily\fontsize{10.000000}{12.000000}\selectfont \(\displaystyle T_{D3}\)}%
\end{pgfscope}%
\begin{pgfscope}%
\definecolor{textcolor}{rgb}{0.000000,0.000000,0.000000}%
\pgfsetstrokecolor{textcolor}%
\pgfsetfillcolor{textcolor}%
\pgftext[x=1.642419in,y=1.103352in,left,base]{\color{textcolor}\rmfamily\fontsize{10.000000}{12.000000}\selectfont \(\displaystyle M_{KK}^2\)}%
\end{pgfscope}%
\begin{pgfscope}%
\definecolor{textcolor}{rgb}{0.000000,0.000000,0.000000}%
\pgfsetstrokecolor{textcolor}%
\pgfsetfillcolor{textcolor}%
\pgftext[x=0.626960in,y=2.818531in,left,base]{\color{textcolor}\rmfamily\fontsize{10.000000}{12.000000}\selectfont \(\displaystyle I\)}%
\end{pgfscope}%
\begin{pgfscope}%
\definecolor{textcolor}{rgb}{0.000000,0.000000,0.000000}%
\pgfsetstrokecolor{textcolor}%
\pgfsetfillcolor{textcolor}%
\pgftext[x=1.922793in,y=2.818531in,left,base]{\color{textcolor}\rmfamily\fontsize{10.000000}{12.000000}\selectfont \(\displaystyle II\)}%
\end{pgfscope}%
\begin{pgfscope}%
\definecolor{textcolor}{rgb}{0.000000,0.000000,0.000000}%
\pgfsetstrokecolor{textcolor}%
\pgfsetfillcolor{textcolor}%
\pgftext[x=3.456307in,y=2.818531in,left,base]{\color{textcolor}\rmfamily\fontsize{10.000000}{12.000000}\selectfont \(\displaystyle III\)}%
\end{pgfscope}%
\end{pgfpicture}%
\makeatother%
\endgroup%
		}
        	\caption{Classical}\label{fig:regionsClassical}
	\end{subfigure}
	\begin{subfigure}[t]{0.49\linewidth}
		\centering
		\resizebox{\linewidth}{!}{
	        	%% Creator: Matplotlib, PGF backend
%%
%% To include the figure in your LaTeX document, write
%%   \input{<filename>.pgf}
%%
%% Make sure the required packages are loaded in your preamble
%%   \usepackage{pgf}
%%
%% Figures using additional raster images can only be included by \input if
%% they are in the same directory as the main LaTeX file. For loading figures
%% from other directories you can use the `import` package
%%   \usepackage{import}
%% and then include the figures with
%%   \import{<path to file>}{<filename>.pgf}
%%
%% Matplotlib used the following preamble
%%
\begingroup%
\makeatletter%
\begin{pgfpicture}%
\pgfpathrectangle{\pgfpointorigin}{\pgfqpoint{4.110671in}{3.074012in}}%
\pgfusepath{use as bounding box, clip}%
\begin{pgfscope}%
\pgfsetbuttcap%
\pgfsetmiterjoin%
\definecolor{currentfill}{rgb}{1.000000,1.000000,1.000000}%
\pgfsetfillcolor{currentfill}%
\pgfsetlinewidth{0.000000pt}%
\definecolor{currentstroke}{rgb}{1.000000,1.000000,1.000000}%
\pgfsetstrokecolor{currentstroke}%
\pgfsetdash{}{0pt}%
\pgfpathmoveto{\pgfqpoint{0.000000in}{0.000000in}}%
\pgfpathlineto{\pgfqpoint{4.110671in}{0.000000in}}%
\pgfpathlineto{\pgfqpoint{4.110671in}{3.074012in}}%
\pgfpathlineto{\pgfqpoint{0.000000in}{3.074012in}}%
\pgfpathclose%
\pgfusepath{fill}%
\end{pgfscope}%
\begin{pgfscope}%
\pgfsetbuttcap%
\pgfsetmiterjoin%
\definecolor{currentfill}{rgb}{1.000000,1.000000,1.000000}%
\pgfsetfillcolor{currentfill}%
\pgfsetlinewidth{0.000000pt}%
\definecolor{currentstroke}{rgb}{0.000000,0.000000,0.000000}%
\pgfsetstrokecolor{currentstroke}%
\pgfsetstrokeopacity{0.000000}%
\pgfsetdash{}{0pt}%
\pgfpathmoveto{\pgfqpoint{0.310278in}{0.279012in}}%
\pgfpathlineto{\pgfqpoint{4.010671in}{0.279012in}}%
\pgfpathlineto{\pgfqpoint{4.010671in}{2.974012in}}%
\pgfpathlineto{\pgfqpoint{0.310278in}{2.974012in}}%
\pgfpathclose%
\pgfusepath{fill}%
\end{pgfscope}%
\begin{pgfscope}%
\definecolor{textcolor}{rgb}{0.000000,0.000000,0.000000}%
\pgfsetstrokecolor{textcolor}%
\pgfsetfillcolor{textcolor}%
\pgftext[x=2.160474in,y=0.223457in,,top]{\color{textcolor}\rmfamily\fontsize{10.000000}{12.000000}\selectfont \(\displaystyle \sigma\)}%
\end{pgfscope}%
\begin{pgfscope}%
\definecolor{textcolor}{rgb}{0.000000,0.000000,0.000000}%
\pgfsetstrokecolor{textcolor}%
\pgfsetfillcolor{textcolor}%
\pgftext[x=0.254723in,y=1.626512in,,bottom,rotate=90.000000]{\color{textcolor}\rmfamily\fontsize{10.000000}{12.000000}\selectfont \(\displaystyle \log(T/M_{Pl}^2)\)}%
\end{pgfscope}%
\begin{pgfscope}%
\pgfpathrectangle{\pgfqpoint{0.310278in}{0.279012in}}{\pgfqpoint{3.700393in}{2.695000in}}%
\pgfusepath{clip}%
\pgfsetrectcap%
\pgfsetroundjoin%
\pgfsetlinewidth{1.505625pt}%
\definecolor{currentstroke}{rgb}{0.000000,0.000000,1.000000}%
\pgfsetstrokecolor{currentstroke}%
\pgfsetdash{}{0pt}%
\pgfpathmoveto{\pgfqpoint{0.310278in}{2.270127in}}%
\pgfpathlineto{\pgfqpoint{0.942305in}{1.938175in}}%
\pgfpathlineto{\pgfqpoint{0.942305in}{1.938175in}}%
\pgfusepath{stroke}%
\end{pgfscope}%
\begin{pgfscope}%
\pgfpathrectangle{\pgfqpoint{0.310278in}{0.279012in}}{\pgfqpoint{3.700393in}{2.695000in}}%
\pgfusepath{clip}%
\pgfsetrectcap%
\pgfsetroundjoin%
\pgfsetlinewidth{1.505625pt}%
\definecolor{currentstroke}{rgb}{0.000000,0.000000,1.000000}%
\pgfsetstrokecolor{currentstroke}%
\pgfsetdash{}{0pt}%
\pgfpathmoveto{\pgfqpoint{0.310278in}{2.391202in}}%
\pgfpathlineto{\pgfqpoint{0.942305in}{2.280552in}}%
\pgfpathlineto{\pgfqpoint{0.942305in}{2.280552in}}%
\pgfusepath{stroke}%
\end{pgfscope}%
\begin{pgfscope}%
\pgfpathrectangle{\pgfqpoint{0.310278in}{0.279012in}}{\pgfqpoint{3.700393in}{2.695000in}}%
\pgfusepath{clip}%
\pgfsetrectcap%
\pgfsetroundjoin%
\pgfsetlinewidth{1.505625pt}%
\definecolor{currentstroke}{rgb}{0.000000,0.000000,1.000000}%
\pgfsetstrokecolor{currentstroke}%
\pgfsetdash{}{0pt}%
\pgfpathmoveto{\pgfqpoint{0.310278in}{1.816398in}}%
\pgfpathlineto{\pgfqpoint{0.942305in}{1.595097in}}%
\pgfpathlineto{\pgfqpoint{0.942305in}{1.595097in}}%
\pgfusepath{stroke}%
\end{pgfscope}%
\begin{pgfscope}%
\pgfpathrectangle{\pgfqpoint{0.310278in}{0.279012in}}{\pgfqpoint{3.700393in}{2.695000in}}%
\pgfusepath{clip}%
\pgfsetrectcap%
\pgfsetroundjoin%
\pgfsetlinewidth{1.505625pt}%
\definecolor{currentstroke}{rgb}{0.000000,0.000000,1.000000}%
\pgfsetstrokecolor{currentstroke}%
\pgfsetdash{}{0pt}%
\pgfpathmoveto{\pgfqpoint{0.310278in}{1.937474in}}%
\pgfpathlineto{\pgfqpoint{0.942305in}{1.937474in}}%
\pgfpathlineto{\pgfqpoint{0.942305in}{1.937474in}}%
\pgfusepath{stroke}%
\end{pgfscope}%
\begin{pgfscope}%
\pgfpathrectangle{\pgfqpoint{0.310278in}{0.279012in}}{\pgfqpoint{3.700393in}{2.695000in}}%
\pgfusepath{clip}%
\pgfsetrectcap%
\pgfsetroundjoin%
\pgfsetlinewidth{1.505625pt}%
\definecolor{currentstroke}{rgb}{0.000000,0.000000,1.000000}%
\pgfsetstrokecolor{currentstroke}%
\pgfsetdash{}{0pt}%
\pgfpathmoveto{\pgfqpoint{0.943642in}{1.937474in}}%
\pgfpathlineto{\pgfqpoint{3.212719in}{0.269012in}}%
\pgfpathlineto{\pgfqpoint{3.212719in}{0.269012in}}%
\pgfusepath{stroke}%
\end{pgfscope}%
\begin{pgfscope}%
\pgfpathrectangle{\pgfqpoint{0.310278in}{0.279012in}}{\pgfqpoint{3.700393in}{2.695000in}}%
\pgfusepath{clip}%
\pgfsetrectcap%
\pgfsetroundjoin%
\pgfsetlinewidth{1.505625pt}%
\definecolor{currentstroke}{rgb}{0.000000,0.000000,1.000000}%
\pgfsetstrokecolor{currentstroke}%
\pgfsetdash{}{0pt}%
\pgfpathmoveto{\pgfqpoint{0.943642in}{2.280318in}}%
\pgfpathlineto{\pgfqpoint{2.901890in}{1.223426in}}%
\pgfpathlineto{\pgfqpoint{2.901890in}{1.223426in}}%
\pgfusepath{stroke}%
\end{pgfscope}%
\begin{pgfscope}%
\pgfpathrectangle{\pgfqpoint{0.310278in}{0.279012in}}{\pgfqpoint{3.700393in}{2.695000in}}%
\pgfusepath{clip}%
\pgfsetrectcap%
\pgfsetroundjoin%
\pgfsetlinewidth{1.505625pt}%
\definecolor{currentstroke}{rgb}{0.000000,0.000000,1.000000}%
\pgfsetstrokecolor{currentstroke}%
\pgfsetdash{}{0pt}%
\pgfpathmoveto{\pgfqpoint{0.943642in}{1.937474in}}%
\pgfpathlineto{\pgfqpoint{4.010527in}{0.809927in}}%
\pgfpathlineto{\pgfqpoint{4.010527in}{0.809927in}}%
\pgfusepath{stroke}%
\end{pgfscope}%
\begin{pgfscope}%
\pgfpathrectangle{\pgfqpoint{0.310278in}{0.279012in}}{\pgfqpoint{3.700393in}{2.695000in}}%
\pgfusepath{clip}%
\pgfsetrectcap%
\pgfsetroundjoin%
\pgfsetlinewidth{1.505625pt}%
\definecolor{currentstroke}{rgb}{0.000000,0.000000,1.000000}%
\pgfsetstrokecolor{currentstroke}%
\pgfsetdash{}{0pt}%
\pgfpathmoveto{\pgfqpoint{0.943642in}{1.594629in}}%
\pgfpathlineto{\pgfqpoint{2.901890in}{0.496700in}}%
\pgfpathlineto{\pgfqpoint{2.901890in}{0.496700in}}%
\pgfusepath{stroke}%
\end{pgfscope}%
\begin{pgfscope}%
\pgfpathrectangle{\pgfqpoint{0.310278in}{0.279012in}}{\pgfqpoint{3.700393in}{2.695000in}}%
\pgfusepath{clip}%
\pgfsetbuttcap%
\pgfsetroundjoin%
\pgfsetlinewidth{1.505625pt}%
\definecolor{currentstroke}{rgb}{0.501961,0.501961,0.501961}%
\pgfsetstrokecolor{currentstroke}%
\pgfsetdash{{4.500000pt}{3.000000pt}}{0.000000pt}%
\pgfpathmoveto{\pgfqpoint{0.943642in}{0.279012in}}%
\pgfpathlineto{\pgfqpoint{0.943642in}{2.974012in}}%
\pgfusepath{stroke}%
\end{pgfscope}%
\begin{pgfscope}%
\pgfpathrectangle{\pgfqpoint{0.310278in}{0.279012in}}{\pgfqpoint{3.700393in}{2.695000in}}%
\pgfusepath{clip}%
\pgfsetbuttcap%
\pgfsetroundjoin%
\pgfsetlinewidth{1.505625pt}%
\definecolor{currentstroke}{rgb}{0.501961,0.501961,0.501961}%
\pgfsetstrokecolor{currentstroke}%
\pgfsetdash{{4.500000pt}{3.000000pt}}{0.000000pt}%
\pgfpathmoveto{\pgfqpoint{2.901944in}{0.279012in}}%
\pgfpathlineto{\pgfqpoint{2.901944in}{2.974012in}}%
\pgfusepath{stroke}%
\end{pgfscope}%
\begin{pgfscope}%
\pgfsetrectcap%
\pgfsetmiterjoin%
\pgfsetlinewidth{0.803000pt}%
\definecolor{currentstroke}{rgb}{0.000000,0.000000,0.000000}%
\pgfsetstrokecolor{currentstroke}%
\pgfsetdash{}{0pt}%
\pgfpathmoveto{\pgfqpoint{0.310278in}{0.279012in}}%
\pgfpathlineto{\pgfqpoint{0.310278in}{2.974012in}}%
\pgfusepath{stroke}%
\end{pgfscope}%
\begin{pgfscope}%
\pgfsetrectcap%
\pgfsetmiterjoin%
\pgfsetlinewidth{0.803000pt}%
\definecolor{currentstroke}{rgb}{0.000000,0.000000,0.000000}%
\pgfsetstrokecolor{currentstroke}%
\pgfsetdash{}{0pt}%
\pgfpathmoveto{\pgfqpoint{0.310278in}{0.279012in}}%
\pgfpathlineto{\pgfqpoint{4.010671in}{0.279012in}}%
\pgfusepath{stroke}%
\end{pgfscope}%
\begin{pgfscope}%
\definecolor{textcolor}{rgb}{0.000000,0.000000,0.000000}%
\pgfsetstrokecolor{textcolor}%
\pgfsetfillcolor{textcolor}%
\pgftext[x=2.234482in,y=1.239327in,left,base]{\color{textcolor}\rmfamily\fontsize{10.000000}{12.000000}\selectfont \(\displaystyle T_{F1}\)}%
\end{pgfscope}%
\begin{pgfscope}%
\definecolor{textcolor}{rgb}{0.000000,0.000000,0.000000}%
\pgfsetstrokecolor{textcolor}%
\pgfsetfillcolor{textcolor}%
\pgftext[x=2.234482in,y=1.627633in,left,base]{\color{textcolor}\rmfamily\fontsize{10.000000}{12.000000}\selectfont \(\displaystyle T_{D3}\)}%
\end{pgfscope}%
\begin{pgfscope}%
\definecolor{textcolor}{rgb}{0.000000,0.000000,0.000000}%
\pgfsetstrokecolor{textcolor}%
\pgfsetfillcolor{textcolor}%
\pgftext[x=2.234482in,y=1.019123in,left,base]{\color{textcolor}\rmfamily\fontsize{10.000000}{12.000000}\selectfont \(\displaystyle T_{D1}\)}%
\end{pgfscope}%
\begin{pgfscope}%
\definecolor{textcolor}{rgb}{0.000000,0.000000,0.000000}%
\pgfsetstrokecolor{textcolor}%
\pgfsetfillcolor{textcolor}%
\pgftext[x=2.234482in,y=0.559261in,left,base]{\color{textcolor}\rmfamily\fontsize{10.000000}{12.000000}\selectfont \(\displaystyle M_{KK}^2\)}%
\end{pgfscope}%
\begin{pgfscope}%
\definecolor{textcolor}{rgb}{0.000000,0.000000,0.000000}%
\pgfsetstrokecolor{textcolor}%
\pgfsetfillcolor{textcolor}%
\pgftext[x=0.626960in,y=2.818531in,left,base]{\color{textcolor}\rmfamily\fontsize{10.000000}{12.000000}\selectfont \(\displaystyle I\)}%
\end{pgfscope}%
\begin{pgfscope}%
\definecolor{textcolor}{rgb}{0.000000,0.000000,0.000000}%
\pgfsetstrokecolor{textcolor}%
\pgfsetfillcolor{textcolor}%
\pgftext[x=2.477156in,y=2.818531in,left,base]{\color{textcolor}\rmfamily\fontsize{10.000000}{12.000000}\selectfont \(\displaystyle II\)}%
\end{pgfscope}%
\end{pgfpicture}%
\makeatother%
\endgroup%
		}
        	\caption{Quantum-corrected}\label{fig:regionsQuantum}
	\end{subfigure}%
	\caption{Tensions of the various 4d strings and the Kaluza--Klein scale
		along the path \eqref{D1limit}, and the three associated regions.
	    Including quantum corrections Region III ceases to exist as the tension of a D3-string
		never goes below the string scale in that case.
	}\label{fig:regions}
\end{figure}
These three regions are illustrated for the classical tensions in figure \ref{fig:regionsClassical}.
We then see that, even if one cannot directly associate a scale to the action
of an instanton, one can relate their magnitude to ratios of string scales. As
we will see in section \ref{ss:QES}, in the present setup this relation can be
easily understood in terms of the c-map to the vector multiplet sector of type
IIA CY compactifications. It would be interesting to see if it can also be
extended to compactifications with lower supersymmetry, along the lines of the
analysis in \cite{Font:2019cxq}.

As follows from the discussion of section \ref{sec:setup}, each of the strings
above is dual to a critical fundamental string, and so their presence will
non-trivially modify the effective field theory
\cite{Lee:2019wij,Lee:2019xtm}. One should therefore consider
separately each of these regions when trying to extract the hypermultiplet
metric from the corrected contact potential \eqref{contact}. Indeed, Region I
can be seen as a classical region of the trajectory in which the instanton
corrections are not very significant, and the string scale is the standard
cutoff scale below which we only have moduli and their Kaluza--Klein
excitations. The contact potential, whose classical piece is derived from a
supergravity action obtained from such string, can be straightforwardly trusted
in such a region. In Region II the fundamental string is no longer the
lightest, being replaced by the D1-string. Having all the D1-string oscillation
modes below the string scale, it is a priori not clear whether the use of the
contact potential is still valid. A more reasonable approach would be to
perform an S-duality transform that exchanges D1- and F1-strings and extract
the metric from the contact potential in this new duality frame. We implement
such a strategy in the next subsection. It turns out that, due to the covariance
of the contact potential, the information obtained in both frames is
essentially the same, and so the approach of \cite{Marchesano:2019ifh} can
still be applied to analyse this region. The same conclusion does not hold,
however, for Region III. In that region not only we have multiple 4d strings
whose tension is below $M_s^2$, it also happens that the lightest of them, the
D1-string, is below the Kaluza--Klein scale. As such the interpretation of the
hypermultiplet metric in terms of the contact potential becomes more subtle, as
we discuss below.

Given the differences between the above regions, it is unlikely that an expression like \eqref{limitchiD1} with a simple cutoff condition of the form \eqref{instantonCondition} will capture the physics in each of them. In the following we will seek a more suitable expression for the contact potential that will allow us to infer the corrections to the hypermultiplet metric in each region separately. To that end, one may consider the initial expression for the corrections given by \eqref{chicorr} and, following \cite{RoblesLlana:2006is,RoblesLlana:2007ae}, split it in two terms:
\begin{align}\label{chicorr-1}
    \chi_{(-1)} & = -\frac{\tau_2^2}{8(2\pi)^3} \,  \chi_E(X) \sum_{(m,n)\in \mathbb{Z}^2\backslash 0}  \frac{1}{|m\tau + n|^3}\, , \\
 \chi_{(1)} & = \frac{\tau_2^2}{8(2\pi)^3} \sum_{\bf{k}>0} n_{\bf{k}}^{(0)} \sum_{(m,n)\in \mathbb{Z}^2\backslash 0}  \frac{1+ 2\pi |m\tau + n|k_at^a}{|m\tau + n|^3}\, e^{-S_{m,n}^{\bf{k}} }\, .
 \label{chicorr1}
\end{align}
Along the D1 limit \eqref{D1limit} only terms with $n=0$ contribute significantly to the sums that define both quantities. For the first one those give
\begin{equation}
	\chi_{(-1)}  \simeq - \frac{\tau_2^{-1}}{4(2\pi)^3} \chi_E(X)  \zeta(3) \,,
\end{equation}
while for the second one finds
\begin{equation}
	\chi_{(1)} \simeq \frac{\tau_2^{-1}}{8(2\pi)^3} \sum_{\bf{k}>0} n_{\bf{k}}^{(0)}  \sum_{m\neq 0}  \frac{1+ 2\pi |m| \tau_2 k_at^a}{|m|^3}\, e^{-2\pi |m| \tau_2 k_at^a }  \simeq - \frac{\tau_2^{-1}}{4(2\pi)^3}  \zeta(3) \sum_{\bf{k>0}}^{\Omega} n_{\bf{k}}^{(0)}\, ,
\end{equation}
using that the series $\sum_{k>0} 1/k^3$ converges quickly to $\zeta(3)$ for the first few terms, and defining $\Omega$ such that $2\pi\tau_2 t^a k_a \ll 1$. We finally arrive at the expression
\begin{equation}\label{newchicorr}
	\chi_{\rm corr}   \simeq  \frac{\tau_2^{-1}}{4(2\pi)^3}  \zeta(3) \left[- \chi_E(X) + \sum_{\bf{k>0}}^{\Omega} n_{\bf{k}}^{(0)}\right] \, ,
\end{equation}
which refines the correction term in \eqref{limitchiD1}, in the sense that it
gives a precise estimate for $k_0^\text{max} \sim \tau_2^{-1}$. Notice that
this corresponds to the value $\eps=0$ in the approximation
\eqref{chargeGrowthD1New} for the D(-1)-charge $k_0$, while for $k_a$ the
bound could be different. Using the results of \cite{Marchesano:2019ifh}
hints to a quantum-corrected trajectory of infinite length. In the next
subsection we will rederive these results in an S-dual frame, and argue that
the infinite distance indeed occurs but because the 4d dilaton varies along
the quantum corrected trajectory. Infinite excursions along K\"ahler
coordinates are, in contrast, obstructed by quantum corrections.

%%%%%%%%%%%%%%%%%%%%%%%%%%%%%%%%%%%%%%%%%%%%%%%%%
%%%%%%%%%%%%%%%%%%%%%%%%%%%%%%%%%%%%%%%%%%%%%%%%%
\subsection{S-duality and the F1 Limit}
%%%%%%%%%%%%%%%%%%%%%%%%%%%%%%%%%%%%%%%%%%%%%%%%%
%%%%%%%%%%%%%%%%%%%%%%%%%%%%%%%%%%%%%%%%%%%%%%%%%

As already pointed out, as soon as $g_s >1$ in the D1 limit we enter Region II,
in which the D(-1)-instanton corrections start to be relevant and D1-strings
become lighter than fundamental strings. Due to the appearance of this lighter
string, the EFT that one derives from the fundamental string breaks down. The
appropriate strategy to derive the hypermultiplet metric would then be to
consider an S-dual frame in which the r\^oles of  D1- and F1-strings have been
interchanged, and therefore the lightest string is the fundamental one. Then,
one may consider the contact potential corresponding to such fundamental string
and derive the hypermultiplet metric from it. Equivalently, one may analyse the
quantum corrections to the contact potential that arises along the F1 limit
\eqref{F1limit}, as we will do in this section.

By construction, the contact potential $\chi$ transforms covariantly under any
$SL(2,\mathbb{Z})$ transformation. In particular under S-duality it transforms
as
\begin{align}\label{Sdu}
    \chi\longrightarrow \chi' = \frac{\chi}{|\tau|}\, ,
\end{align}
where as in section \ref{sec:setup} we will use primes to denote quantities
in the F1 limit.  As a result, the quantum corrections contributing to $\chi$
are reshuffled and, e.g., the actions of worldsheet and D1-brane instantons
are exchanged with one another: 
\begin{equation}
		S^\text{WS}_{\bf{k}}(t') = 
		S^\text{D1}_{\bf{k}}(\tau,t)\, .
\end{equation}
It is easy to check that in the F1 limit the dominating corrections to the
contact potential within \eqref{chicorr} will come from $\chi_{\rm pert}$ and
$\chi_{\rm WS}$, and so it can be approximated by 
\begin{eqn}\label{chiF1}
	\chi' \sim &~\frac{1}{12}(\tau_2')^2\mathcal{K}_{abc}{t'}^a{t'}^b{t'}^c - \frac{\chi_E(X)}{4(2\pi)^3} \zeta(3) (\tau_2')^2 \\
	&+\frac{(\tau_2')^2}{4(2\pi)^3}\sum_{\bf{k} > 0} n_{\bf{k}}^{(0)}\text{Re}\left[\text{Li}_3\left(e^{-2\pi k_a {t'}^a}\right)+2\pi k_a {t'}^a \text{Li}_2\left(e^{-2\pi  k_a {t'}^a}\right)\right]\, .
\end{eqn}
Then, using that at small arguments polylogarithms behave as 
\begin{align}
	\text{Li}_s(x)\sim x\quad\text{as}\quad x\to0\,,
\end{align}
the dominating part of the contact potential is found to be 
\begin{align}\label{limitchiF1}
	\chi'\sim \frac{1}{12}(\tau_2')^2 \mathcal{K}_{abc}{t'}^a{t'}^b{t'}^c  + \frac{(\tau_2')^2}{32 \pi^3 } \zeta(3)   \left[- \chi_E(X) + \sum_{\bf{k>0}}^{\Omega} n_{\bf{k}}^{(0)}\right] \, .
\end{align}
Taking into account the transformation \eqref{Sdu}, this reproduces the
previous result \eqref{newchicorr}. It follows that in the F1 limit
perturbative $\alpha'$ corrections play the r\^ole that D(-1) effects had in
the D1 limit while, as stated before, D1-brane instantons are replaced by
worldsheet instantons. In the region in which the latter are subdominant, the
corrected 4d dilaton behaves as
\begin{align}\label{chiF1fullapprox}
    \chi' = \text{const.}\, (\tau_2')^2 \sim e^{3\sigma}\,. 
\end{align}
By inserting this result in the metric \eqref{metricD1}, one sees that the 4d
dilaton direction provides a constant contribution to the norm of the velocity
vector along the trajectory \eqref{F1limit}. In the following we would like to
argue that this behaviour still holds when we proceed along the trajectory deep
into Region II and beyond, with the constant in \eqref{chiF1fullapprox} being a
minimum quantum-corrected volume after which the K\"ahler coordinates cannot
decrease any further. In this sense, quantum corrections would heavily deform
the classical trajectory \eqref{F1limit}, preventing it to advance in the
K\"ahler directions and inducing a variation of constant velocity along the 4d
dilaton direction. 

In order to support this claim let us consider a related setup, the quantum
K\"ahler moduli space $\mathcal{M}_{QKM}$ describing the vector multiplet
sector in type IIA string theory on a Calabi--Yau $X$, see \cite{Hori:2003ic}.
Such a moduli space, recently considered in \cite{Lee:2019wij} in the context
of infinite distance limits, is related to the one probed in the F1 limit by a
standard c-map.  As it is well known, mirror symmetry maps $\mathcal{M}_{QKM}$
to the classical complex structure moduli space $\mathcal{M}_{cs}$ of the
mirror Calabi--Yau $Y$. The mirror map is defined near the large complex
structure point, which can be chosen to be at $y=0\in \mathcal{M}_{cs}$. Close
to this point the periods of the holomorphic three-form $\Omega_Y$,
\begin{align}
	X^A = \int_{\gamma_A} \Omega_Y \,,\qquad F_B = \int_{\gamma^B}\Omega_Y\,, \qquad A,B=0,\dots h^{2,1}(Y)\,,
\end{align}
where $(\gamma_A, \gamma^B)$ is a symplectic basis of three-cycles on $Y$, take
the form
\begin{align}
    X^0 = 1 +\mathcal{O}(y^a)\,,\qquad X^a = \frac{1}{2\pi i}\log(y^a) +\mathcal{O}(y^a)\,,\qquad a=1,\dots h^{2,1}(Y)\,.
\end{align}
These can be used to define flat coordinates of $\mathcal{M}_{cs}$, that near
the large complex structure point $y=0$ take the form
\begin{align}\label{zaya}
    z^a(y^a)=\frac{X^a}{X^0} = \frac{1}{2\pi i}\log(y^a) + \mathcal{O}(y^a)\,.
\end{align}
Close to the point $y=0$, the logarithmic term in \eqref{zaya} dominates and we
can identify $z^a$ with the classical complexified K\"ahler moduli of $X$ via
the classical mirror map $z^a = \frac{1}{2\pi i} {\rm log}\, y^a$. Away from
$y=0$ this na\"ive map will however start to fail, and it is the variables
$z^a$, extended by analytic continuation, that must be considered as the
coordinates  of the full quantum K\"ahler moduli space of $X$. For instance,
approaching a small volume point which na\"ively would correspond to the  limit
$y^a\rightarrow 1$, one sees that the approximation $z=\frac{1}{2\pi i} {\rm
log}\, y^a$ is no longer valid. As a consequence of the polynomial term,
instead of going to the limit where $t^a=\text{Im}\,z^a \rightarrow 0$, the
K\"ahler coordinates get frozen to a constant value, and in the limit where the
classical volume of the Calabi--Yau shrinks to zero, one ends up with a limit
where the quantum volume is just a constant. For specific examples such as the quintic and other models with few K\"ahler moduli, this behaviour has been explicitly verified in the context of the SDC in \cite{Blumenhagen:2018nts,Joshi:2019nzi}.

Via the c-map, one may embed the above IIA vector moduli space as a totally
geodesic submanifold of $\mathcal{M}_{HM}$ of type IIB on the same Calabi--Yau,
and then apply the above observations to the trajectory along the F1 limit.
One sees that it is precisely when the worldsheet instantons should be as
relevant as the perturbative $\alpha'$ corrections, or equivalently when the
string scale goes below the Kaluza--Klein scale, that the K\"ahler moduli used to
define the trajectory are no longer flat coordinates of the quantum moduli
space. Using the appropriate coordinates one sees that the classical ones stop
at a constant value:
\begin{equation}
	{t'}^a \stackrel{~\eqref{F1limit}~}{\longrightarrow} {t'}_{QM}^a = \text{const}\,.
\end{equation}
Nevertheless, the trajectory in moduli space does not stop, as it continues to
progress at a constant rate along the 4d dilaton coordinate towards weaker and
weaker coupling. 

Going back to the original trajectory describing the D1 limit, this implies
that quantum corrections shield the classical trajectory \eqref{D1limit} from
entering Region III in figure \ref{fig:regionsQuantum}, while instead inducing a
geodesic variation along the 4d dilaton coordinate.  In particular, this
picture allows one to evaluate the sum over the D1-instantons charges in the D1
limit. For this, notice that since the quantum K\"ahler moduli in the F1 limit
are constant, the worldsheet instanton actions are as well: 
\begin{align}
    \text{Re}(S_{WS})\stackrel{~\eqref{F1limit}~}{\longrightarrow} \text{const.}
\end{align}
Now, the D1-instanton actions in the dual D1 limit \eqref{D1limit} are interchanged with those of worldsheet instanton actions via S-duality
\begin{align}
    \text{Re}(S_{D1}) = \tau_2 k_a t^a = 
    k_a (t')^a =\text{const.}
\end{align}
Thus, as one moves along the trajectory \eqref{D1limit}, only the tower of D(-1)-instantons acquire an asymptotically vanishing action, while the D1-instanton actions asymptote to constant values. Notice that the action of the D(-1)-instantons itself is just given by the 10d dilaton. As argued in Appendix \ref{app:metric} the 10d dilaton does not receive quantum corrections itself, allowing the D(-1)-instanton action to asymptotically vanish.

%%%%%%%%%%%%%%%%%%%%%%%%%%%%%%%%%%%%%%%%%%%%%%%%%
%%%%%%%%%%%%%%%%%%%%%%%%%%%%%%%%%%%%%%%%%%%%%%%%%
\subsection{Quantum Emergent Strings}\label{ss:QES}
%%%%%%%%%%%%%%%%%%%%%%%%%%%%%%%%%%%%%%%%%%%%%%%%%
%%%%%%%%%%%%%%%%%%%%%%%%%%%%%%%%%%%%%%%%%%%%%%%%%

The emerging picture from the analysis of this section so far can be summarised as follows, see also figure \ref{fig:regions}. As we cross from Region I to Region II in the D1 limit, D(-1)-instantons start getting dominant and we should change the duality frame to consider the S-dual F1 limit, in which the fundamental string is the lightest. There the quantum corrections due to the D(-1)-instantons are mapped to $\alpha'$ perturbative corrections, which become stronger as we proceed towards smaller volumes in the F1 limit. Eventually perturbative and non-perturbative $\alpha'$ corrections prevent the quantum volume of the Calabi--Yau to vanish, obstructing the trajectory to enter Region III. Instead, they induce a constant velocity variation along the 4d dilaton coordinate, sourcing an infinite distance geodesic towards weak coupling. 

According to the Emergent String Conjecture, as we proceed along this trajectory there should be a unique string that becomes tensionless, which is dual to either a type II or a heterotic fundamental string. To see how this happens, let us consider the lightest string in the D1 limit. The tension of such a D1-string, including the effects of quantum corrections is given by
\begin{align}
    \frac{T_{D1}}{M_\text{Pl}^2} = \frac{T_{D1}}{M_s^2}\frac{M_s^2}{M_\text{Pl}^2} = \frac{T_{D1}}{M_s^2} \frac{1}{\chi}\,. 
\end{align}
with $\chi$ the quantum-corrected 4d dilaton \eqref{chicorr}. Through \eqref{newchicorr}, one can infer that it grows as $\chi\sim e^{3 \sigma/2}$ in the far end of the trajectory. What remains to be determined is the scaling of the string tension in string units. Classically, it is given by
\begin{align}\label{TD1TF1}
 \frac{T_{D1}}{M_s^2} = \tau_2 \sim e^{-\frac{3}{2}\sigma}\,
\end{align}
that is by the variation of the 10d dilaton. By S-duality, the same relation should hold after quantum corrections. Finally, as argued in Appendix \ref{app:metric} the 10d dilaton, seen as a coordinate in $\mathcal{M}_{HM}$, is not affected by quantum corrections. As a result \eqref{TD1TF1} should hold along the whole trajectory, and we obtain a tensionless D1-string in the limit \eqref{D1limit} with tension
\begin{align}
    \frac{T_{D1}}{M_\text{Pl}^2} \sim e^{-3\sigma}\,. \label{TD1D1}
\end{align}

One may apply the same reasoning to see how the tension of other strings scale
in the quantum-corrected regime. For instance, the tension of a D3-string
wrapped on some curve $\gamma_a\subset X$ behaves as: 
\begin{align}\label{TD3D1}
	\frac{T_{D3}}{M_\text{Pl}^2} \sim \begin{cases} e^{-\sigma/2}\,,& \text{classically}\,,\\ \chi^{-1}\,, &\text{after quantum corrections}\,,\end{cases}
\end{align}
reflecting the fact that the tension of the D3-string in string units scales
like the action of the D1-instanton wrapped on the same curve. As a consequence
the quantum corrections increase the speed at which the D3-string becomes
tensionless to $\chi^{-1} \sim e^{-3\sigma/2}$, but not as much as they do for
the D1-string. Similar results hold for the other strings that can be obtained
by wrapping branes on suitable cycles in $X$. Thus, all tension of
4d strings, including the tension of the F1-string which scales
like $M_s^2/M_\text{Pl}^2\sim e^{-3\sigma/2}$, are parametrically higher than
the tension of the D1-string, which thus remains to be the lightest string also
in the heavily quantum-corrected regime. 

Via S-duality, the D1-string maps to the fundamental string probing an internal
manifold of constant volume due to the quantum corrections. Thus, we see that
in the D1 limit we find a tensionless string that is dual to a weakly-coupled
fundamental type IIB string with tension
\begin{align}\label{TF1F1}
    \frac{T'_{F1}}{M_\text{Pl}^2} \sim e^{-3\sigma}\,,
\end{align}
realising the Emergent String Conjecture in this infinite distance limit. In
some sense this shows that the classical limit \eqref{D1limit} that one
initially wanted to reach does not exist as part of the quantum-corrected
moduli space, and that is diverted to a limit in which a weakly-coupled,
tensionless fundamental string is obtained. Notice that the endpoint of the
trajectory is at finite distance in K\"ahler moduli space, since the quantum
volume asymptotes to a constant. However, since the trajectory proceeds with
constant velocity towards a weak 4d coupling limit, $\chi\rightarrow \infty$,
the length of the trajectory remains infinite. As we comment in Appendix
\ref{app:metric}, this behaviour seems to be reflected by the metric components of the
K\"ahler moduli, for which the infinite distance behaviour is removed when
taking into account quantum corrections. On the other hand, the quantum-corrected metric
component corresponding to $\tau_2$ reproduces the classical scaling and
therefore yields the infinite part to the distance in moduli space. 

To complete the above picture, one has to ensure that the D1 limit does not yield a decompactification towards a higher-dimensional theory before the lightest string becomes tensionless. Approximating the Kaluza--Klein scale as
\begin{align}
    \frac{M_{\rm KK}^2}{M_\text{Pl}^2} \simeq \frac{1}{\mathcal{V}(X)^{1/3}} \frac{M_s^2}{M_\text{Pl}^2}\, ,
\end{align}
and deducing the quantum-corrected scaling of the K\"ahler moduli via S-duality
\begin{align}
    t^a =\tau_2' t'^a_\text{QM} \sim e^{3/2\sigma}\,, 
\end{align}
one obtains that the mass of KK modes scales as 
\begin{align}
    \frac{M_{\rm KK}^2}{M_\text{Pl}^2} \simeq e^{-3\sigma}\,.
\end{align}
Notice that this matches the scaling of the tension of the D1-string, and so both scales are comparable in the quantum-corrected regime.\footnote{That is, the trajectory is obstructed to cross to Region III in figure \ref{fig:regions} exactly when $T_{\rm D1}^{1/2} \sim M_{\rm KK}$.}
However, since the tower of string states is denser than the KK tower, this limit is equi-dimensional \cite{Lee:2019wij}. Similar results can be obtained from a direct analysis of the F1 limit.

\subsubsection*{The c-map and Vector Multiplets}

It is quite instructive to compare the above trajectories in hypermultiplet moduli space to analogous ones that one could follow in vector multiplet moduli space. The simplest way to do so is to apply the c-map construction \cite{Cecotti:1988qn,Ferrara:1989ik}, which in our case involves compactifying type IIB on $\IR^{1,2} \times S^1 \times X$ and performing a T-duality on the $S^1$. Under this operation, the type IIB hypermultiplet moduli space $\mathcal{M}_{\rm HM}^{\rm IIB}$ is mapped to the dimensional reduction of type IIA vector multiplet $\mathcal{M}_{\rm VM}^{\rm IIA}$ on a circle, which is the twisted hypermultiplet moduli space  $\mathcal{M}_{\rm THM}^{\rm IIA}$. In the following we will apply such construction along the trajectory of the D1 limit \eqref{D1limit}, since an analogous path in $\mathcal{M}_{\rm VM}^{\rm IIA}$ has been analysed in the context of the Swampland Distance \cite{Corvilain:2018lgw} and Emergent String \cite{Lee:2019wij} conjectures. 

One may trivially embed the classical trajectory \eqref{D1limit} in the moduli space of $S^1 \times X$, with $S^1$ a circle whose radius in string units $L_B = R_B M_s$ remains constant along the trajectory. Upon T-duality on this circle we obtain type IIA on $S^1 \times X$, such that \cite{Seiberg:1996ns}
\begin{subequations}
\label{cmap}
\begin{align}
\label{LA}
& L_A \equiv R_A M_s = \frac{1}{L_B}\, ,\\
\label{g4A}
&  g_{4, A}^{-1} = g_{4, B}^{-1} {L_B} = R_B M_{\text{Pl}, B}\, ,\\
\label{RAMA}
&  R_A M_{\text{Pl}, A} =  g_{4, A}^{-1} L_A = g_{4, B}^{-1} \, ,
\end{align}
\end{subequations}
where $R_A$ is the $S^1$ radius in the type IIA side and $g_4$, $M_{\text{Pl}}$ stand for the 4d dilaton and 4d Planck mass in either the type IIA or IIB side of the duality. After the T-duality the K\"ahler moduli of $X$ and the 10d dilaton follow the same trajectory \eqref{D1limit} as they did in the type IIB side, but this now forms part of the twisted hypermultiplet moduli space $\mathcal{M}_{\rm THM}^{\rm IIA}$. Notice that classically all the quantities in \eqref{cmap} are constant. As such one can also interpret this trajectory as a classical path of infinite distance in $\mathcal{M}_{\rm VM}^{\rm IIA}$. 

One interesting feature of this duality is that it maps pairs of instantons and strings in $\IR^{1,2} \times S^1$ to particles in $\IR^{1,2} \times S^1$. Indeed, type IIB D(-1)-instantons and D1-strings wrapped on the $S^1$ are mapped to D0-branes on the type IIA side, respectively wrapping and not wrapping the $S^1$. The same statement applies to D1-instantons and D3-branes wrapping a two-cycle $\g^a$ of $X$, which map to type IIA D2-branes wrapping the same cycle. Because both elements of a pair go to the same object in type IIA, namely a particle, it follows the tension of the string $T$ and the instanton action $S$ must be related as $T = M_s^2 \re S$. This gives a rationale as to why certain non-perturbative effects become important as soon as the tension of a string is comparable with that of a fundamental string. 

On the type IIB side and from a four-dimensional viewpoint, proceeding along the trajectory \eqref{D1limit} results in several strings lowering their
tension with respect to $M_\text{Pl}^2$. From a three-dimensional perspective,
these strings will be seen as particles when they wrap the $S^1$. The mass of
such particles with respect to the 3d Planck mass reads
\begin{equation}
	M_\text{F1} = M_s^2 R_B = \frac{M_{\text{Pl},3}}{\chi} \, ,\quad \quad M_\text{D1} = T_\text{D1} R_B = \frac{M_{\text{Pl},3}}{\tau_2\chi} \, ,
\end{equation}
for the lightest winding modes of the fundamental and D1-brane, respectively.
Notice that these relations are a direct consequence of \eqref{Msp}, which was
interpreted in terms of a species scale. In the present setup, the r\^ole of
species scale is played by the winding mode scales of the F1- and D1-strings,
respectively, and the fact that these scales decrease exponentially along the
infinite distance trajectory signals that one approaches a decompactification
limit, which tensionless but codimension one strings cannot prevent. This is
even more manifest after T-duality to type IIA, where $M_\text{F1} = 1/R_A$ and
$M_\text{D1} \raw M_\text{D0} =1/R_{11}$ are related to the Kaluza--Klein scales of the
c-map and M-theory radii, respectively. While the M-theory radius grows even at
the classical level, quantum corrections imply that $R_A M_{\text{Pl},3}$ also
grows exponentially, although at a slower rate. Above these two scales one
finds M-theory compactified on $T^2 \times X$, and the quantum corrections
leading to $\chi$ can be computed along the lines of \cite{Collinucci:2009nv}.
Finally, it is easy to see that by reducing M-theory to type IIA along $R_A$
and performing a c-map along $R_{11}$ one again recovers type IIB on $S^1
\times X$ but now with a trajectory that corresponds to the F1 limit
\eqref{F1limit}. We thus observe a tight relation between limits of tensionless
strings and decompactification limits. 

In fact, our results are very suggestive when compared to the observations made
in \cite{Lee:2019wij} on the type IIA vector multiplet moduli space. Indeed,
since the D(-1)-instantons are c-mapped to D0-branes, taking their corrections
into account in $\mathcal{M}_{\rm HM}^{\rm IIB}$ can be interpreted as
integrating-in D0 branes in $\mathcal{M}_{\rm VM}^{\rm IIA}$. As discussed
above, the analogous of the D1 limit in $\mathcal{M}_{\rm VM}^{\rm IIA}$ is
the strong coupling/M-theory limit in which D0-branes become massless.
Following \cite{Lee:2019wij}, to appropriately analyse this setup one should
consider M-theory in 5d, which from a four-dimensional type IIA perspective is
at infinite distance since one has to blow up the radius of the M-theory
circle. Then, having a constant 5d Planck mass forces the M-theory K\"ahler
moduli to be constant. Therefore, in this analogous setup of the D1 limit the
infinite distance along the K\"ahler directions is also removed due to the
inclusion of quantum corrections, while the infinite distance coming from
scaling the dilaton is still present. The difference is that while we see a
tensionless fundamental string in $\mathcal{M}_{\rm HM}^{\rm IIB}$, the
analogous limit in $\mathcal{M}_{\rm VM}^{\rm IIA}$ yields a
decompactification.

%%%%%%%%%%%%%%%%%%%%%%%%%%%%%%%%%%%%%%%%%%%%%%%%%
%%%%%%%%%%%%%%%%%%%%%%%%%%%%%%%%%%%%%%%%%%%%%%%%%
\subsection{Vanishing Fibre Limit}\label{ss:D3limit}
%%%%%%%%%%%%%%%%%%%%%%%%%%%%%%%%%%%%%%%%%%%%%%%%%
%%%%%%%%%%%%%%%%%%%%%%%%%%%%%%%%%%%%%%%%%%%%%%%%%
We would now like to extend the above discussion to the third limit established
in the classical hypermultiplet moduli space, the D3 limit \eqref{D3limit}.
Since we know that the quantum corrected D1 limit is dual to the weak-coupling
limit of type IIB string theory, we only have to establish the duality between
the D3 limit and the D1 limit at quantum level to show that there is a weakly-coupled, tensionless fundamental string theory that describes the asymptotics
of the D3 limit. As explained in section \ref{sec:setup}, the D3 limit
\eqref{D3limit} can be reached starting from the D1 limit in cases where the
compactification manifold is an elliptically fibred Calabi--Yau threefold by
performing two T-dualities along the fibre or, more precisely, a particular
Fourier--Mukai transform.

However, in the new frame the relevant quantum corrections do not come anymore
solely from D1/D(-1)-instantons and the procedure used so far to incorporate
quantum effects on the hypermultiplet moduli space has to be adapted. To
achieve this, we must first consider which class of instantons we expect to
contribute significantly to the metric. As Fourier--Mukai transforms map
instantons into instantons, we are assured that we only need to consider the
transformation of the relevant instanton actions.\footnote{Via the c-map, this
can be seen by formulating BPS particles in the type IIA side in a categorical
language, where the Fourier--Mukai transform is an equivalence
\cite{Alexandrov:2013yva,Andreas:2004uf}.} In the case at hand these are found
in the new frame through the transformation of the moduli \eqref{FMTransform}:
\begin{equation}\label{instantonactionD3}
	\text{Re}(S^{D}_{k_\Lambda})=\tau_2|k_\Lambda z^\Lambda| \longrightarrow
	\tau_2 \left|\left(k_1 +\frac{1}{2}k_\alpha K^\alpha\right)+ k_0 z^1 + k_\alpha \left(z^1z^\alpha +\frac{1}{2}K^\alpha (z^1)^2\right) \right|\,,
\end{equation}
where $k_0$ stands for the D(-1)-charge in the original duality frame, and we
singled out the D1-instanton charge along the elliptic fibre, $k_1$. After the
transformation, the first term is associated to a D(-1)-instanton with charge
$q_0 = k_1+\frac{1}{2}k_\alpha K^\alpha$, while the second is the action an instanton of
charge $q_1 = k_0$ coming from a D1-brane wrapping the elliptic fibre. Finally, the last term
in \eqref{instantonactionD3} has precisely the form of an instanton of charge $q^\alpha$ associated
to D3-branes wrapping a vertical divisor:
\begin{equation}
	\text{Re}(S_{D3|_{D_\alpha}}) = q^\alpha \,\text{Vol}(D_\alpha)\,,\qquad 
	\text{Vol}(D_\alpha)=\eta_{\alpha\beta}\left(t^1t^\beta+\frac{1}{2}K^\beta(t^1)\right)\,,
\end{equation}
where $\eta_{\alpha\beta}= D_\alpha\cdot D_\beta$ denotes the intersection pairing of vertical divisors. Generalising the notation used in the definition of the D1/D(-1)-instanton action, we can write the RHS of \eqref{instantonactionD3} as
\begin{gather}
	\text{Re}({S^D_{k_A}}) =\tau_2\, k_A z^A\,,\\
	k_A = (k_0,k_1,k_\alpha,0)\,,\qquad
	z^A = (1,z^1,z^\alpha, \text{Vol}(D_a) )\, .
\end{gather}
In this language, the Fourier--Mukai transformation acts, as expected, on the lattice of charges:
\begin{equation}\label{D3instantonFMtransformation}
	\text{Re}({S^D_{k^A}})\longrightarrow \text{Re}({S^D_{q^A})=\tau_2|q_Az^A|}\,.
\end{equation}
The charges after the transformation are related to the original D1/D(-1)-charges $k_A$ by matrix multiplication (cf. \cite{Corvilain:2018lgw}):
\begin{equation}
	q_A = 
	\begin{pmatrix}
		k_1+\frac{1}{2}k_\alpha K^\alpha\\k_1\\0\\\eta^{\alpha\beta}k_\beta
	\end{pmatrix}
	=
	\begin{pmatrix}
		0&-1&\frac{1}{2}K^\alpha&0\\
		1&0&0&\frac{1}{2}K_\alpha\\
		0&0&0&-\eta_{\alpha\beta}\\
		0&0&\eta^{\alpha\beta}&0
	\end{pmatrix}
	k_A\,,
\end{equation}
where $\eta^{\alpha\beta}$ is the inverse of the intersection pairing and $K_\alpha = \eta_{\alpha\beta}K^\alpha$. We
conclude that D1/D(-1)-instantons are transformed into D3/D1/D(-1)-instantons
that have a D1-charge only along the fibre. In the case of a trivial fibration,
the Chern class of the base vanishes and we obtain the expected behaviour under
a double T-duality, and the presence of D3-instantons thus accounts for the
non-trivial curvature of the base under the transform \eqref{FMTransform}.

Unfortunately, one cannot directly use the contact potential \eqref{chisplit} to
take into account quantum corrections, as it does not include contributions
from D3-brane instantons. However, as reviewed in Appendix
\ref{app:D3-instantons}, corrections to the 4d dilaton are known
for generic, mutually local D-brane instantons:
\begin{align}\label{4ddilatonD3}
	\chi =& 
	\frac{1}{6}\tau_2^2 \mathcal{K}_{abc}{t}^a{t}^b{t}^c
	-\frac{\chi(X)}{192 \pi}\
	+\frac{\tau_2}{8\pi^2}\sum_{q} \Omega(q)\sum_{m=1}^\infty \frac{\text{Re}(S^D_{q})}{m}K_1(2\pi m \text{Re}(S^D_{q}))\,.
\end{align}
The sum is taken over all D-instantons of charge $q$, and $\Omega$ denotes the
associated Donaldson--Thomas invariant. In the case of D1/D(-1)-instantons, the
latter reduce to either the Euler density or Gopakumar--Vafa invariants and one
recovers equation \eqref{chiD}. 

The D3 limit differs from the original D1 limit only by the Donaldson--Thomas
invariants, as the asymptotics of the classical instanton action are the same,
see equation \eqref{D3instantonFMtransformation}. As already alluded to
earlier, in the present case, the Fourier--Mukai transform corresponds to a
monodromy action $M$ on the K\"ahler moduli space \cite{Cota:2019cjx}. We can
therefore use that Donaldson--Thomas invariants do not change under
monodromies, i.e. $\Omega(Mk)=\Omega(k)$ to conclude that $\chi$ in the D1
and D3 limit coincide:
\begin{align}\label{chiD3approx}
	\chi'' \sim \chi_\text{class.} + \frac{\tau_2^{-1}}{4(2\pi)^3}  \zeta(3) \left[- \chi_E(X) + \sum_{\bf{k>0}}^{\Omega} n_{\bf{k}}^{(0)}\right] \sim \chi \,,
\end{align}
with the second term behaving as the quantity in brackets in \eqref{newchicorr}.
The only difference is the interpretation: after T-dualising, $k_0$ 
counts the instanton charge of a D1-instanton wrapped along the fibre rather
than a D(-1)-instanton in the original limit. 

Note that one needs to ensure that other types of instantons, such as those
coming from NS5- and D5-branes, are also parametrically smaller. The classical scaling of
the actions for the instantons in all the limits keeping the Planck mass
constant are given in table \ref{tab:actions}. As one can see, D5-branes do not
acquire a vanishing action in either frames. A similar statement also
holds for NS5-instantons. Furthermore, perturbative corrections are missing in
table \ref{tab:actions}, since there is no obvious way how to express them in
terms of instanton actions. As noted in the previous subsection, in the
F1 limit they take over the r\^ole of the D(-1)-instantons in the D1 limit. 

\begin{table}[!t]
	\centering
	\begin{tabular}{|c|c|c|c|c|c|c|c|c|c|}
		\hline 
		Limit & D(-1) & D1$|_{\mathcal{E}}$ &D1$|_{\mathcal{C}}$ & D3$|_{D_0}$ & D3$|_{D_\alpha}$& D5$|_X$ & NS5$|_X$ & WS$|_{\mathcal{E}}$ &WS$|_{\mathcal{C}}$ \\\hline \hline
		D1, eq. \eqref{D1limit}& $e^{-\frac{3}{2}\sigma}$&$e^{-\frac{1}{2}\sigma}$& $e^{-\frac{1}{2}\sigma}$ & $e^{\frac{1}{2}\sigma}$&$e^{\frac{1}{2}\sigma}$&$e^{\frac{3}{2}\sigma}$& $e^{0}$ &$e^\sigma$&$e^\sigma$\\ \hline 
		F1, eq. \eqref{F1limit} &$e^{\frac{3}{2}\sigma}$&$e^\sigma$& $e^\sigma$ & $e^{\frac{1}{2}\sigma}$ &$e^{\frac{1}{2}\sigma}$&$e^{0}$& $e^{\frac{3}{2}\sigma}$&$e^{-\frac{1}{2}\sigma}$&$e^{-\frac{1}{2}\sigma}$\\\hline 
		D3, eq. \eqref{D3limit}&$e^{-\frac{1}{2}\sigma}$&$e^{-\frac{3}{2}\sigma}$& $e^{\frac{1}{2}\sigma}$ & $e^{-\frac{3}{2}\sigma}$ &$e^{-\frac{1}{2}\sigma}$&$e^{\frac{1}{2}\sigma}$& $e^{0}$ &$e^{-\sigma}$& $e^{\sigma}$\\ \hline   
	     
	\end{tabular}
	\caption{Asymptotic scaling of the real part of the action for the different instantons in each limit. $\mathcal{C}\subset B_2$ denotes a curve in the base.}
	\label{tab:actions}
\end{table}

Again, the fact that $\chi''$ is not constant tells us that the fundamental
string scale in the new limit changes in terms of the Planck mass. We saw that
classically the string coming from a D3-string is the lightest with classical
tension 
\begin{align}
    \frac{T''_{D3}}{M_\text{Pl}^2}=\tau_2'' (t^1)'' \frac{1}{\chi''_{\rm cl}}\,. 
\end{align}
One may however fear that this string no longer becomes tensionless the fastest
after quantum corrections. Notice that in the present setup quantum corrections
could obstruct the limit in which the area of the fibre vanishes, analogously
to the obstruction of the small-volume point in the F1 limit, or the correction
to the D3-brane tension in the D1 limit, see \eqref{TD3D1}.

We can answer this question via duality to the D1 limit, by using that the
D3-string tension in string units is mapped under the Fourier--Mukai transform
to $\tau_2$
\begin{align}
    \tau_2''  (t^1)'' \sim \tau_2\,.
\end{align}
As discussed in Appendix \ref{app:metric}, the scaling of $\tau_2$ in the D1
limit is not affected by quantum corrections. Therefore, the tension of the
D3-brane wrapped on the fibre should not be affected by the quantum corrections
relevant in the D3 limit. In other words, the D3/D1/D(-1)-instantons above do not affect the vanishing of the fibre volume in this limit. 
Taking into account the corrected expression for $\chi$ in \eqref{chiD3approx}, we
obtain that the tension of the D3-strings in the D3 limit behaves as
\begin{align}
    \frac{T''_{D3}}{M_\text{Pl}^2} \sim e^{-3\sig}\,,
\end{align}
which matches the corrected tension of the D1-string in the D1 limit and thus
by extension the corrected tension of the F1 string in the F1 limit. 

We thus conclude that after taking into account instanton corrections, we can
effectively describe the vanishing-fibre limit in moduli space by a
tensionless, weakly-coupled type IIB string theory.

%%%%%%%%%%%%%%%%%%%%%%%%%%%%%%%%%%%%%%%%%%%%%%%%%
%%%%%%%%%%%%%%%%%%%%%%%%%%%%%%%%%%%%%%%%%%%%%%%%%

%%%%%%%%%%%%%%%%%%%%%%%%%%%%%%%%%%%%%%%%%%%%%%%%%
%%%%%%%%%%%%%%%%%%%%%%%%%%%%%%%%%%%%%%%%%%%%%%%%%

%%%%%%%%%%%%%%%%%%%%%%%%%%%%%%%%%%%%%%%%%%%%%%%%%
%%%%%%%%%%%%%%%%%%%%%%%%%%%%%%%%%%%%%%%%%%%%%%%%%

\section{Infinite Distance Limits in Type I String Theory}\label{sec:typeI}

We have so far investigated how quantum corrections alter the geometry around
classical infinite distance points in the hypermultiplet sector of type IIB
string theory. We showed that taking these corrections into account, a
tensionless string arises in three infinite distance limits, where quantum
effects are under control, which is dual to a tensionless and weakly-coupled
fundamental string of type IIB string theory. Moreover these strings already
arose classically and instantons only enhanced this behaviour.

In this section, we would like to address yet another string theory setup which has
infinite distance limits and in which instanton corrections play an important
r\^ole, namely four-dimensional $\mathcal{N}=2$ compactifications of type I
string theory. 

Type I string theory is well known to be related to type IIB string theory by
modding out worldsheet parity, arriving to a theory of unoriented strings where
half of the supersymmetry gets broken. To obtain a four-dimensional
$\mathcal{N}=2$ EFT, we are therefore led to consider a type I string theory
compactified on 
\begin{equation}
	X = K3\times T^2\,.
\end{equation}
To make contact with the setup analysed in the D3 limit  we can can
interpret this as a compactification on a trivial elliptic fibration over a $K3$
base. While somewhat similar, there are important differences between the two.
Indeed, the volume of the torus, its complex structure and the dilaton now give
rise to vector multiplets. Accordingly, it is now the vector multiplet moduli
space that receives quantum corrections in the form of instantons, rather than
the hypermultiplet sector. Finally, as a consequence of the mixing of the complex
structure and K\"ahler sector, it is a priori harder to classify points at infinite distance in terms of mixed Hodge structures \cite{Grimm:2018ohb}. 

For definiteness, the following analysis will focus on the BSGP orbifold
\cite{Bianchi:1990tb,Gimon:1996rq} which can be thought of as the
$T^4/\mathbb{Z}_2\times T^2$ orbifold limit of $K3\times T^2$. The reason for
this choice is that all relevant quantum contributions are known in this case
\cite{Camara:2008zk}. 

This model has an unbroken $U(16)$ gauge group realised
by D9-branes, while the D5-gauge group gets broken due to the Green--Schwarz
mechanism. In the following, we will ignore Wilson line moduli such that the
part of the vector multiplet moduli space that we are interested in is spanned
by three physical vector multiplet moduli:
\begin{align}\label{TypeImoduli}
	S=e^{-\phi}\, T\, \mathcal{V}+i\,b \,,\qquad S'=e^{-\phi} \,T +i\,b'\,,\qquad U\,.
\end{align}
The complexified 4d dilaton is denoted by $S$; $U$ is the complex
structure of the torus; $T=\text{Vol}(T^2) M_s^2$ and
$\mathcal{V}=\text{Vol}(K3) M_s^4$ are respectively the volumes of the torus
and the K3, measured in units of the string scale $M_s$, which type I string
theory inherits from type IIB string theory. Finally $e^{-\phi}$ is the
10d type I dilaton and the two axions, $b$ and $b'$ are respectively obtained by
dualising the type I two-form or reducing it along the torus. 

In terms of the physical fields, the Planck mass is given by:
\begin{equation}\label{typeIMPL}
	M_\text{Pl}^2 = e^{-\phi}\, \text{Re}(S) M_s^2 \,.
\end{equation}
From the emergent string proposal \cite{Lee:2019wij} and the type
IIB examples discussed previously, we would like to check whether tensionless
four-dimensionless strings emerge at points of infinite distance. Obvious candidates
for such strings are type I D1-branes in Minkowski spacetime,
whose tension is given by 
\begin{equation}
	T_\text{D1} = e^{-\phi}M_s^2\,.
\end{equation}
From the definition of the Planck mass, we see that the real part of $S$
controls the D1-brane tension in Planck units. 
%%%%%%%%%%%%%%%%%%%%%%%%%%%%%%%%%%%%%%%%%%%%%%%%%
%%%%%%%%%%%%%%%%%%%%%%%%%%%%%%%%%%%%%%%%%%%%%%%%%
\subsection{Small Torus Limit}
%%%%%%%%%%%%%%%%%%%%%%%%%%%%%%%%%%%%%%%%%%%%%%%%%
%%%%%%%%%%%%%%%%%%%%%%%%%%%%%%%%%%%%%%%%%%%%%%%%%
Having introduced the moduli space for type I string theory compactified on
$K3\times T^2$, we now want to look for classical infinite distance limits in
this moduli space. We focus on limits that keep $S$ constant, and one possible
case for the type I vector multiplet moduli \eqref{TypeImoduli} that we
can consider is
\begin{align}\label{TypeILimit}
 S =\text{const.}\,,\qquad \re\, S'\sim e^{-\sigma}\,, \qquad U=\text{const.}\,,\qquad b=b'=0\,. 
\end{align}
Since $S'$ is related to the volume of the torus, we can think of this limit
as a small torus limit and taking $K3\times T^2$ as a trivial elliptic
fibration, we can interpret it as the type I analogue of the D3 limit in type
IIB hypermultiplet moduli space. Unlike the previous type IIB hypermultiplet
analysis we can obtain the vector multiplet geometry directly from a K\"ahler
potential, which classically reads
\begin{align}
	K_\text{cl}=-\log(S+\bar S)-\log(S'+\bar S')-\log(U+\bar U)\,,
\end{align}
In the limit \eqref{TypeILimit}, the only relevant component of the classical metric is
 given by:
\begin{align}
   g_{S'\bar S'}=\frac{1}{(S'+\bar S')^2}\sim e^{2\sigma}\,, 
\end{align}
such that the associated line element gives indeed rise to an
infinite distance. 

Classically, this limit keeps the ratio $T_\text{D1}/M_\text{Pl}^2$ constant,
and the type I string does not become tensionless in Planck units. On the other
hand, D1-branes wrapping one of the torus one-cycles are becoming massless in
this classical limit, giving rise to a KK-like tower 
\begin{align}\label{D1p}
	\frac{M_\text{D1}^2}{M_{\rm Pl}^2}\sim n^2 \frac{\re\, S'}{\re\, S} \sim n^2 e^{-\sigma}\,.
\end{align}
We would thus argue that the limit \eqref{TypeILimit} is a decompactification limit. 
Still, these are classical statements and the intuition we earned from the
previous sections tells us that we should take into account quantum corrections
as we consider a limit for which a cycle shrinks to zero size. From our previous experience, we expect
either perturbative or instanton corrections to dominate.

For the BSGP orbifold, the relevant corrections to the K\"ahler potential have
been computed in \cite{Camara:2008zk} and take the following form:
\begin{equation} \label{Kcorrected}
	K=-\log\left(S+\bar S + V_\text{1-loop} + V_\text{D1} \right)
	-\log(S'+\bar S')-\log(U+\bar U)\,.
\end{equation}
As one can see the only relevant corrections are to the 4d dilaton.
Similarly to type IIB hypermultiplets, these are of both perturbative and
non-perturbative origin, the latter corresponding to D1-instantons wrapping
the torus:
\begin{subequations}\label{typeIKahlerCorrections}
\begin{align}
	V_\text{1-loop} &= -\frac{4\pi}{6}\frac{E(iU,2)}{S'+\bar S'}\, ,\\
	V_\text{D1} &= -\frac{1}{2\pi}\sum\limits_{k>j\ge 0\,,p>0}\frac{e^{-2\pi k p S'}}{(kp)^2}\left(\frac{\hat{A}_K(\mathcal{U})}{(S'+\bar S')}-\frac{2ikp}{\mathcal{U}-\bar{\mathcal{U}}}\frac{E_{10}(\mathcal{U})}{\eta^{24}(\mathcal{U})}\right)\, .
\end{align}
\end{subequations}
Let us shortly review the properties of these corrections. On the one hand, the perturbative
contributions $V_\text{1-loop}$ are phrased in terms of non-holomorphic
Eisenstein series of order $k$:
\begin{align}
        E(U,k)\equiv \frac{1}{\zeta(2k)}\sum\limits_{(j_1,j_2)\ne (0,0)}\frac{(\text{Im}\,U)^k}{|j_1+j_2U|^{2k}}\,. 
\end{align}
On the other hand, the D1-instanton corrections $V_\text{D1}$ can be expressed in terms of quasi-modular forms related to the usual Eisenstein series $E_{2k}$, depending on complex structure of the instanton worldsheet wrapping the torus:
\begin{align}
	\hat{\mathcal{A}}_K(\mathcal{U})=\frac{1}{12 \eta^{24}}\left(\hat E_2 E_4 E_6 +2 E_6^2+3E_4^3\right)\,,
	\qquad \mathcal{U}=\frac{j+ipU}{k}\,.
\end{align}
Note that $\hat{\mathcal{A}}_K$ is only quasi-modular due to the appearance of $\hat E_2(\tau) = E_2(\tau)-\frac{3}{\pi\text{Im}(\tau)}$. 

In the limit $\text{Re}(S')\rightarrow 0$ both perturbative and instanton corrections become important, dominating over the classical $S+\bar S$ term of the K\"ahler potential. In particular:
\begin{equation}\label{Vcorrscaling}
	V_\text{1-loop} + V_\text{D1} \sim \frac{1}{S'+\bar{S}'}\sim e^{\sigma}\,.
\end{equation}
By $\mathcal{N}=2$ supersymmetry, these corrections induce at the same time a modification of the gauge kinetic function of the $U(16)$ open string gauge group \cite{Camara:2008zk}:
\begin{align}\label{corrfU16}
    f_{U(16)}=S+&\frac{\pi S'}{2}-12\log \eta(iU) -\frac{1}2{\sum\limits_{k>j\ge 0 \,,p>0}\frac{e^{-2\pi k p S'}}{(kp)}}{\mathcal{A}}_f(\mathcal{U})\,, 
\end{align}
where 
\begin{align}\label{Af}
	{\mathcal{A}}_f=\frac{1}{12\eta^{24}} \left(E_2 E_4 E_6 -\frac{5}{12}E_6^2-\frac{7}{12}E_4^3\right)\,, 
\end{align}
and $\eta$ denotes the Dedekind function.

In the regime \eqref{TypeImoduli}, the gauge kinetic function also gets
dominated by instanton effects, and depends on the precise value the variable
$U$ through the modular function $\mathcal{A}_f(\mathcal{U})$. For cases where
the torus complex structure is large, $\text{Re}(U)\gg1$, the imaginary part of
$\mathcal{U}$ will generically also be large, such that we can use the
approximation
\begin{align}
	\mathcal{A}_f(\mathcal{U})=-\frac{49}{2}+\mathcal{O}\left(e^{2\pi i\mathcal{U}}\right)\,.
\end{align}
The sum over instanton numbers in \eqref{corrfU16} will yield a diverging
contribution in the limit $S'\rightarrow 0$ where the instanton action vanishes
and thus the exponential suppression is lifted. We can make the rough estimate
\begin{align}
        f_{U(16)}&\sim S+\frac{49}{4}{\sum\limits_{k>j\ge 0\,,p>0}\frac{e^{-2\pi k p S'}}{(kp)}} \sim S+\frac{49}{4}\sum\limits_{p>0}\frac{1}{2\pi p^2 S'}\,,
\end{align}
where we used that the summand does not explicitly depend on $j$ and imposed
$k<\left(2\pi p S'\right)^{-1}$ in order for the exponential to be
$\mathcal{O}(1)$.  We thus conclude that in the limit $S'\rightarrow 0$ the
gauge kinetic function of the open string gauge group blows up, yielding a
weakly-coupled gauge theory.

Along the trajectory \eqref{TypeILimit}, we can track the tensions of BPS
strings by looking at the decay constants of the axions they couple to
magnetically. For instance, the D1-string couples magnetically to Im$S$, and
its tension is given in Planck units by the decay constant $f_{\text{Im}S}$:
\begin{align}
    \frac{T_{D1}}{M_{\rm Pl}^2}=f_{\text{Im}S} =\sqrt{\partial_S\partial_{\bar S} K} \, .
\end{align}
Classically, the tension of the D1-string is constant since 
\begin{align}
    \partial_S\partial_{\bar S} K_\text{cl} = \frac{1}{(S+\bar S)^2}\,,
\end{align}
but as the K\"ahler potential receives quantum corrections, we find that the corrected D1-tension scales as
\begin{eqn}\label{td1quantum}
	\left.\frac{T_\text{D1}}{M_\text{Pl}^2}\right|_\text{corr} &=\sqrt{ \partial_S\partial_{\bar S} K_\text{corr}}
	= \left(\text{Re}(S_\text{eff})\right)^{-1}
	\sim \text{Re}(S')\sim e^{-\sigma}\,,\\
	\text{Re}(S_\text{eff}) &= \frac{1}{2}\left( S+\bar{S}+ V_\text{1-loop} + V_\text{D1} \right)\,,
\end{eqn}
where we used the approximation \eqref{Vcorrscaling}. Consequently the type I
string in fact becomes tensionless after taking into account instanton
corrections. This is in stark contrast with the behaviour of the limits we
studied in type IIB. There, quantum corrections were merely making the strings
become tensionless at a faster rate. In the present case however, they are
strong enough to completely overwhelm the classically finite tension and drive
it to zero.

Still, we have to check whether there are no other 4d strings
becoming tensionless in this limit at the same rate as the D1-string or faster.
Candidates for these strings arise from D5-branes wrapped on the entire K3 or
wrapped on the torus times a 2-cycle inside K3. The latter string couples
magnetically to axions that are part of the hypermultiplet moduli space and
their tension is hence not affected by our limit in vector multiplet moduli
space. The string coming from the D5-brane wrapped on $K3$ on the other hand
couples magnetically to $\text{Im}\,S'$. Hence its tension scales like 
\begin{align}
    \frac{T_{D5|_{K3}}}{M_{\rm Pl}^2}=f_{\text{Im}\,S'} =\frac{1}{\sqrt{(S'+\bar S')}}\rightarrow \infty\,,
\end{align}
such that the string becomes infinitely heavy asymptotically. The only
tensionless string is thus the D1-string as type I string theory does not have
any D3- or D7-branes.

Similarly, as we are going to a regime where the D1-string becomes very light,
the type I supergravity description might not be valid anymore. Using the
philosophy of the previous sections, we should instead look at the effective
theory that can be derived from the tensionless string: the D1-string itself.
As it is well known, the type I setup studied so far is S-dual to the heterotic
string on $K3\times T^2$ and the classical limit \eqref{TypeILimit} is
translated to the heterotic limit 
\begin{align}\label{clashet}
    \text{Re}\,S_H = \text{const.}\,,\qquad \re\, T_{H}\sim e^{-\sigma}\,,\qquad U=\text{const.}\,,
\end{align}
where $S_H$ is the heterotic 4d dilaton, $T_H$ the volume of the
torus in units of the heterotic string scale $M_{H}$ and $U$, as before, the torus
complex structure. The heterotic string on $K3\times T^2$ receives quantum
corrections from winding and momentum states along the internal
directions. Their effect is analogous to the type I effect studied above,\footnote{In fact, type I corrections in \cite{Camara:2008zk}
have been calculated via duality to the heterotic string.} such that the corrected
trajectory in the heterotic frame is given by:
\begin{align}\label{hetlimit}
    \text{Re}\,S_H|_\text{corr.} \sim e^{\sigma} \,,\qquad  \re\, T_{H}\sim e^{-\sigma}\,,\qquad U=\text{const.}\,,
\end{align}
and heterotic string becomes weakly coupled. As a consequence, the heterotic
string scale $M_H$ vanishes asymptotically when measured in Planck units: 
\begin{align}
    \frac{M_H^2}{M_{\rm Pl}^2} = (\text{Re}\,S_H)^{-1}\rightarrow e^{-\sigma}\,,
\end{align}
in agreement the D1-string becoming tensionless in the dual Type I limit. 

Still, the question remains of whether the heterotic string tension is comparable with the lowest KK scale (in which case we would indeed have an emergent string limit) or if its tension is above some KK scale, yielding a decompactification limit. The lowest KK-like tower in this setup, with masses given by \eqref{D1p}, corresponds here to heterotic winding modes. We see that, simply using the new scalings of the fields in the vector multiplet \eqref{hetlimit}, we obtain:
\begin{align}\label{mwindclas}
  \frac{M_\text{winding}^2}{M_{\rm Pl}^2}\sim \frac{\re\, T_H}{\re\, S_H} \sim e^{-2\sigma}\,. 
\end{align}
As this is parametrically below the heterotic string scale, we are still in a decompactification regime even after taking into account quantum corrections. The effect of the quantum corrections would then just be to obstruct the small torus limit $T_H\rightarrow 0$ at finite coupling by pushing us towards a weak-coupling limit. This is in agreement with instanton corrections breaking the $SL(2,\mathbb{Z})_{T_H}$ part of the modular group. However, it seems to contradict the results obtained in a dual setup, as we now discuss.

\subsubsection*{Comparison to Dual Setup}
The vector multiplet moduli space of $\mathcal{N}=2$ supergravity theories has already been investigated in other setups, namely that of Calabi--Yau three-folds in type IIA \cite{Corvilain:2018lgw,Lee:2019wij}. Heterotic string theory on $K3\times T^2$ is dual to type IIA on a K3-fibred Calabi--Yau threefold $Z:
K3 \rightarrow \mathbb{P}^1$ which at the same time admits an elliptic
fibration over a Hirzebruch surface $Z:T^2 \rightarrow F_n$
\cite{GarciaEtxebarria:2012zm}. A concrete realisation is
$W\mathbb{P}_{1,1,2,8,12}(24)$, which has $h^{(1,1)}=3$ and therefore gives
rise to three vector multiplets that can be identified with $S, T_H$ and $U$
\cite{GarciaEtxebarria:2012zm}. The three complexified K\"ahler moduli
$z_{a}\,, a=1,2,3$ describe respectively the volume of the elliptic fibre, the
volume of the $\mathbb{P}^1$-fibre of $F_n$, and the volume of the base
$\mathbb{P}^1$. The explicit identification with $S, T_H, U$ is given by:
\begin{align}
    z_1=U\,,\qquad z_2=T_H-U\,, \qquad z_3=S_H-\left(1-\frac{n}{2}\right)U-\frac{n}{2}T_H\,. 
\end{align}
Moreover, the volume of the K3 fibre, which can be thought of as the restriction
of the elliptic fibration to the fibre $\mathbb{P}^1$ is 
\begin{align}
    \mathcal{V}(K3)_{IIA}=UT_H\,. 
\end{align}
We thus see that the heterotic limit $T_H\rightarrow 0$ corresponds to the limit
where the K3 fibre volume vanishes. Our analysis shows that the type I/heterotic limit
is heavily corrected by quantum effects, just as it happens in \cite{Lee:2019wij}.

However, the result of \cite{Lee:2019wij} implies that the quantum obstructions to the limit of vanishing K3 volume give rise to an emergent tensionless heterotic string. In fact a mirror symmetry calculation reveals that the volume of the K3-fibre $\mathcal{V}(K3)_{IIA}$ cannot shrink to values smaller than one in type IIA string units. The point that is reached by this trajectory then corresponds to the conifold point. Compared to our heterotic/type I setup this indicates that $T_H$ should also have a minimal value and therefore should be constant in the quantum-corrected limit. This result seems in direct contradiction with the quantum-corrected limit \eqref{hetlimit}. 

The analogue of the conifold point of type IIA in our heterotic/type I setup is the point of self-dual radius of the torus where one $U(1)$ gauge group enhances to $SU(2)$ due to winding modes that become massless. One could therefore wonder whether in the above discussion we missed effects due to the 1-loop singularity at the gauge enhancement point or due to other sources, such that \eqref{hetlimit} is not the actual limit after taking all quantum corrections into account. Otherwise, our result would mean that one can proceed through this conifold point in the type IIA setup, towards a decompactification limit.

\subsection{Large Complex Structure}
As a second possible limit, we consider the limit of large complex structure of
the $T^2$, i.e. 
\begin{align}\label{Utoinfty}
    \text{Re}\, U \sim e^\sigma \rightarrow \infty\,,
\end{align}
with $\text{Re}\,S$ and $\text{Re}\,S'$ kept fixed. Note first that our
conventions in this section follow \cite{Camara:2008zk} which can differ from
other definitions of the complex structure by a factor of $i$; second, the
large and small complex structure limits are equivalent due to the presence of
an unbroken $SL(2,\mathbb{Z})_U$ modular symmetry. 

Applying the same reasoning as in the previous section, the D1-string and the
D5-brane wrapped on $K3$ have tension 
\begin{align}
	\frac{T_{D1}}{M_{\rm Pl}^2}=(\text{Re}\,S)^{-1}=\text{const.}\,,\qquad \frac{T_{D5}|_{K3}}{M_{\rm Pl}^2}=(\text{Re}\,{S'\;})^{-1}=\text{const.}
\end{align}
and thus no tensionless strings appear, as the tension of D5-branes wrapped on
$T^2$ times a 2-cycle in K3 is again controlled by a modulus in the
hypermultiplet sector. On the other hand, one can notice that D1-strings
wrapped along one of the one-cycles of the torus become massless in this limit,
as classically their mass scales as 
\begin{align}
    \frac{M_{D1}}{M_{\text{Pl}}}\sim \frac{q_1 + Uq_2}{\sqrt{U+\bar U}}\,,
\end{align}
where $q_1$ and $q_2$ denote charges in a suitable basis of the two one-cycles
of the torus. We thus see that as long as $q_2=0$, the D1-particles become
massless:
\begin{align}
	\frac{M_{D1}|_{q_2=0}}{M_{\text{Pl}}}\sim \frac{q_1}{\sqrt{U+\bar U}}\sim e^{-\sigma/2}\,.
\end{align}
To get a better intuition of the situation, let us perform a T-duality along
one of the cycles. Doing so, the type I theory gets mapped to type I' theory,
D1-branes getting mapped to D0-branes localised on D8-branes, and the complex
structure of the torus exchanged with the K\"ahler volume.  The precise duality
map is not relevant for the following discussion, the important point being that
the Buscher rules tell us that the type I' string coupling has to blow up:
\begin{align}
    e^{-\phi^{I'}} \rightarrow e^{-\sigma/2}\,. 
\end{align}
Accordingly, the D0-branes then become massless:
\begin{align}
	\frac{M_{D0}}{M_{\rm Pl}} =e^{-\phi^{I'}} \rightarrow e^{-\sigma/2}\,,
\end{align}
as expected from T-duality. The D0-branes are, however, nothing but the
Kaluza--Klein modes of Ho\v{r}ava--Witten theory, i.e. the strong coupling limit of
type I' theory compactified on $K3\times T^2$ times an extra circle. 

We might ask whether the limit \eqref{Utoinfty} is also heavily affected by
quantum corrections. At the level of the K\"ahler potential we see that the
corrections \eqref{typeIKahlerCorrections} due to the scaling of
$\text{Re}\,U\rightarrow \infty$ are either suppressed, or at most constants
exponentially suppressed by $\text{Re}\,S'$. We thus do not expect the effect
of quantum corrections on the K\"ahler potential---and by extension the string
tensions---to be as significant as in the limit analysed in the previous section.
The gauge kinetic function, however, is again corrected and potentially blows
up. A similar effect should be observable in the T-dual setup. The effects of
quantum corrections to the gauge kinetic functions in type I'/Ho\v{r}ava--Witten
theory have already been studied in great detail in \cite{Gonzalo:2018guu} and
will not be repeated here.

\subsubsection*{Comparison to Type IIA}
Again we can compare the result for the large/small complex structure limit with the IIA dual introduced in the previous subsection. The limits of large/small complex structure $U$ on the type
I side map, via the duality transform outlined above, to $z_1\rightarrow
0/\infty$ in the type IIA geometry. Due to the identification of $z_1$ with the
volume of the elliptic fibre, this corresponds to a $T^2$ limit in the language
of \cite{Lee:2019wij}. The fact that the $SL(2,\mathbb{Z})_U$ modular symmetry
is unbroken reflects that in type IIA the points $z_1\rightarrow
0$ and $z_1\rightarrow \infty$ are the same and related by a monodromy, the
Fourier--Mukai transform. Due to this monodromy the type IIA prepotential is a modular function of $z^1$ \cite{Schimannek:2019ijf} consistent with quantum corrections on the type I side arranging in modular functions of $U$.

Finally, we have seen that the $U\rightarrow \infty$ limit is dual to a
decompactification limit of type I' string theory to Ho\v{r}ava--Witten theory
in one dimension higher. This result is in agreement with the $T^2$-type limits
in type IIA Calabi--Yau compactifications, which were shown to yield a
decompactification to M-theory in one dimension higher \cite{Lee:2019wij}.
%
%%%%%%%%%%%%%%%%%%%%%%%%%%%%%%%%%%%%%%%%%%%%%%%%
%%%%%%%%%%%%%%%%%%%%%%%%%%%%%%%%%%%%%%%%%%%%%%%%%
%%%%%%%%%%%%%%%%%%%%%%%%%%%%%%%%%%%%%%%%%%%%%%%%%

%%%%%%%%%%%%%%%%%%%%%%%%%%%%%%%%%%%%%%%%%%%%%%%%%
%%%%%%%%%%%%%%%%%%%%%%%%%%%%%%%%%%%%%%%%%%%%%%%%%

\section{Conclusions}\label{sec:conclu}

In this work we have tested the Emergent String Conjecture \cite{Lee:2019wij}
in regimes where quantum corrections significantly modify the moduli space
metric. We have done so in two different classes of 4d ${\cal N}=2$
compactifications. We have first considered infinite distance limits along
the hypermultiplet moduli space of type IIB Calabi--Yau compactifications, a
setup already analysed in \cite{Marchesano:2019ifh}. Second, we have
considered trajectories in the vector multiplet moduli space of type I string
theory on $K3 \times T^2$, whose infinite distance limits do not fall into
the classifications made in \cite{Grimm:2018ohb,Lee:2019wij}. 

As pointed out in \cite{Marchesano:2019ifh}, classically the hypermultiplet
moduli space of type IIB on a CY has two types of infinite distance
trajectories. The first one corresponds to moving along the 4d dilaton
towards a weak-coupling limit, and the second type consists of moving in
K\"ahler moduli space keeping the (classical) 4d dilaton constant. Via the
c-map, this second type of trajectories can be related to the ones in vector
multiplet moduli space of type IIA on the same CY, analysed in
\cite{Corvilain:2018lgw,Lee:2019wij}. The main difference is that, while in
type IIA a tower of particles asymptotes to zero mass, in type IIB a tower of
instantons and 4d strings tend to zero action and tension, respectively. This
behaviour for the strings was noticed in \cite{Marchesano:2019ifh}, but its
significance was not investigated. Here, in light of the recent Emergent
String Conjecture, we have taken a direct interest on identifying these light
strings and in unveiling what their presence implies. In particular, we have
labelled each of the limits of section \ref{sec:setup} in terms of its
lightest string, see figure \ref{fig:moduliSpace}. At the end of the day, we
find that this lightest string tension signals the breakdown of the 4d
effective field theory, and that it corresponds to the emerging string of the
proposal made in \cite{Lee:2019wij}.

The emerging picture from our analysis can be described as follows. As we proceed along an infinite distance trajectory of the second type, D-brane instanton corrections start to dominate over the classical contribution to the moduli space metric. The relevance of each instanton tower can be estimated by comparing the tension of the D-brane 4d strings with that of the fundamental string. This relation between string tensions has been depicted in figure \ref{fig:regions} for the D1 limit, which was also analysed in \cite{Marchesano:2019ifh}. There, as soon as the D1-string tension goes below the F1-string tension, namely when crossing to Region II, D(-1)-instanton corrections dominate over the classical moduli space metric. Their effect implies that the 4d dilaton is no longer constant, but instead is forced to flow towards a weak-coupling limit. Quantum effects become even more dramatic when the 4d tension of D3-brane wrapping 2-cycles of the CY become comparable to the F1-string tension, i.e. when crossing to Region III. In there, D1-brane instanton effects should start to dominate. However, via S-duality to the F1 limit and mirror symmetry, we find an obstruction for their action to vanish. In terms of the analysis in \cite{Marchesano:2019ifh} and in Appendix \ref{app:metric}, this can be understood as a quantum obstruction for the K\"ahler coordinates to contribute to an infinite distance trajectory. However, due to quantum corrections the trajectory also proceeds along the 4d dilaton coordinate. A refinement of the analysis in \cite{Marchesano:2019ifh} shows that the metric in this direction remains essentially unaffected by quantum corrections. As a result the trajectory ends up being of infinite distance and, while classically it started as a trajectory with constant 4d dilaton, it ends as a trajectory approaching weak coupling exponentially fast. This weak-coupling limit is maintained when we turn to a dual frame in which the lightest string is nothing but the fundamental type IIB string, realising the Emergent String Conjecture. 

This picture is not only realised in the D1 limit analysed in section \ref{ss:D1limit}, but also in the limits of vanishing fibre in which the lightest string is a D3-brane wrapping a 2-cycle of the CY, as discussed in section \ref{ss:D3limit}. Still, we only analysed a small subset of all possible limits in the type IIB hypermultiplet sector. It would therefore be important to classify all possible limits and to check whether the observed pattern of emerging strings holds in general. In \cite{Lee:2019wij} all possible infinite distance limits in the type IIA vector multiplet moduli space, subject to a constant string-to-Planck mass ratio, have been classified. Interestingly, when applied to the type IIA CY vector multiplet moduli space, each of the limits considered in this work corresponds to a decompactification limit. As the next step one should apply the known of the classification of infinite distance in the type IIA vector multiplet sector to the hypermultiplet moduli space of type IIB CY compactifications, to see if the decompactification limit in vector multiplet moduli space corresponding to the vanishing fibre limit for a $T^4$-fibrations also turns into an emergent string limit when applied to the hypermultiplet sector, as we have observed for our limits. On the other hand if the Calabi--Yau is K3-fibred an emergent heterotic string appears in the IIA vector multiplet moduli space.  It would then be very interesting to understand the physics of this limit in the IIB hypermultiplet analogue.

Instanton corrections also play a non-trivial r\^ole along infinite
distance trajectories in the vector multiplet moduli space of type I
string theory on $K3\times T^2$. For trajectories along which the
torus shrinks while the D9-brane gauge coupling is classically kept
constant, D1-instantons wrapped on the torus together with perturbative
corrections become dominant over the classical contribution to the
K\"ahler potential. These instantons obstruct the finite coupling
limit, in the sense that they bend the classical trajectory towards a
weak-coupling limit in which the D1-string becomes asymptotically
tensionless. In other words, to arrive to the point of vanishing $T^2$ area
 quantum corrections force us to go to vanishing coupling.
This reflects that the classical $SL(2,\mathbb{Z})$
symmetry, that exchanges large and small torus areas, is broken at
finite coupling. On the other hand limits of
large/small complex structure are less sensitive to the D1-instantons
and can be reached without the D1-string becoming tensionless. 

We have found that, naively, each of these limits correspond to a
decompactification limit. On the one hand in the case of the complex structure
limits this is to be expected, because they are dual to decompactification
limits to Ho\v{r}ava-Witten theory. On the other hand, the small torus limit is
supposed to be dual to a K3-fibre vanishing volume limit of the kind analysed
in \cite{Lee:2019wij}. It was there obtained  that this is an equi-dimensional
limit in which an heterotic string emerges, due to a quantum obstruction for
the K3 fibre to reach a vanishing volume. We may directly analyse this dual
heterotic setup by simply applying S-duality to type I on $K3\times T^2$. In
this frame the K3 volume obstruction would translate into an obstruction for a
vanishing $T^2$ area in string units, which would then imply an emerging string
limit. As there is a priori no evident mechanism for such an obstruction, we
find a na\"ive contradiction with the type IIA result. It would be interesting
to carry a more detailed analysis that sheds more light into this puzzle, and
in particular to describe the dual map between the type I and type IIA moduli
spaces at the quantum level.  In this context one should essentially understand
whether the map itself receives quantum corrections or if the corrections
calculated in \cite{Camara:2008zk} already capture all the quantum corrections
to the type IIA quantum K\"ahler moduli space.

Our analysis showed the importance of quantum corrections for the Emergent
String Conjecture in the type IIB hypermultiplet moduli space, and it would be
interesting to extend our analysis to dual setups like for the hypermultiplet
sector of type IIA Calabi--Yau compactifications, see e.g.
\cite{Grimm:2019wtx}. It would be particularly interesting to study limits for
which the type IIB mirror dual is not in a geometric phase, as they might give
new insights into the Emergent String Conjecture and its relation to instanton
corrections to the asymptotic moduli space geometry.

Finally, it would be important to investigate the effect of instanton corrections in $\mathcal{N}=1$ theories, where they not only correct the field space metric but they also induce contributions to the superpotential. Quantum corrections to the $\mathcal{N}=1$ superpotential and their effect for the SDC have been investigated in \cite{Gonzalo:2018guu}. Along these lines one could consider quantum corrections in $\mathcal{N}=1$ setups---both for the superpotential and K\"ahler potential---and confront their effects with the Emergent String Conjecture. 

\bigskip

\bigskip

\vspace*{.05cm}

\centerline{\bf  Acknowledgments}

\vspace*{.05cm}

\bigskip
We would like to thank P. Corvilain, M. Fuchs, D. Kl\"awer, S.-J. Lee, W. Lerche, E. Palti, T. Schimannek and T. Weigand for useful discussions and correspondence. This work is supported by the Spanish Research Agency (Agencia Estatal de Investigaci\'on) through the grant IFT Centro de Excelencia Severo Ochoa SEV-2016-0597, and by the grant PGC2018-095976-B-C21 from MCIU/AEI/FEDER, UE. MW received funding from the European Union's Horizon 2020 research and innovation programme under the Marie Sklodowska-Curie grant agreement No. 713673. The work of MW also received the support of a fellowship from "la Caixa" Foundation (ID 100010434) with fellowship code LCF/BQ/DI18/11660033. MW thanks the CERN Theoretical Physics Department for hospitality.

\newpage

%%%%%%%%%%%%%%%%%%%%%%%%%%%%%%%%%%%%%%%%%%%%%%%%%
%%%%%%%%%%%%%%%%%%%%%%%%%%%%%%%%%%%%%%%%%%%%%%%%%
%%%%%%%%%%%%%%%%%%%%%%%%%%%%%%%%%%%%%%%%%%%%%%%%%
%%%%%%%%%%%%%%%%%%%%%%%%%%%%%%%%%%%%%%%%%%x%%%%%%%
\appendix
%%%%%%%%%%%%%%%%%%%%%%%%%%%%%%%%%%%%%%%%%%%%%%%%%
%%%%%%%%%%%%%%%%%%%%%%%%%%%%%%%%%%%%%%%%%%%%%%%%%
%%%%%%%%%%%%%%%%%%%%%%%%%%%%%%%%%%%%%%%%%%%%%%%%%
%%%%%%%%%%%%%%%%%%%%%%%%%%%%%%%%%%%%%%%%%%%%%%%%%

%%%%%%%%%%%%%%%%%%%%%%%%%%%%%%%%%%%%%%%%%%%%%%%%%
\section{Planck Mass, String Scale, and Duality Frames}\label{app:PlanckMass}
%%%%%%%%%%%%%%%%%%%%%%%%%%%%%%%%%%%%%%%%%%%%%%%%%
%%%%%%%%%%%%%%%%%%%%%%%%%%%%%%%%%%%%%%%%%%%%%%%%%
%%%%%%%%%%%%%%%%%%%%%%%%%%%%%%%%%%%%%%%%%%%%%%%%%

We review here the relation between the Planck mass and the string scale, how
it behaves in the three frames discussed in section \ref{sec:setup} and
\ref{sec:QuantumCorrections}, and the relations between them.  In terms of
string frame variables, the Planck mass is given by
\begin{equation}
	M^2_\text{Pl} = 2\pi \tau_2^2\,\mathcal{V}(t)\, M_s^2\,,\qquad 
	\mathcal{V}(t) = \frac{1}{6} \mathcal{K}_{abc}t^at^bt^c\,,
\end{equation}
where $\mathcal{V}(t)$ is the volume of the compact manifold $X$ expressed in
terms of the triple-intersection numbers $\mathcal{K}_{abc}$ and
$\tau_2=e^{-\phi}$ is the 10d dilaton. In the entirety of this
work, we want the Planck-mass-to-string-scale ratio
$\frac{M^2_\text{Pl}}{M^2_s}$ to remain constant. However, when performing a
duality transform, its interpretation might change, which means that if we want
to compare a given quantity, such as the rate at which a string becomes
tensionless, one needs to keep track of this change. The purpose of this
appendix is to provide some details that supplement the main text. 

\paragraph{S-duality:} here, the transform is a strong-weak duality and the
dilaton transforms as:
\begin{equation}
	\tau \longrightarrow \tau' = -\frac{1}{\tau}\,,\qquad
	\tau_2 \longrightarrow \tau_2' = \frac{\tau_2}{|\tau|^2}\,, 
\end{equation}
and should not have any effect on \emph{dimensionful} K\"ahler moduli,
\begin{equation}
	Z^a = \frac{z^a}{M_s^2} \longrightarrow {Z'}^a = Z^a\,,\qquad
	z^a = b^a +it^a\,.
\end{equation}
To define the dimensionless moduli in the dual frame, one must then make use of
both string scales,
\begin{equation}
	{z'}^a = {Z'}^a (M_s')^2 = z^a \left( \frac{M_s'}{M_s} \right)^2\,,
\end{equation}
such that the Planck mass in terms of the transformed coordinates is given by 
\begin{equation}
	\frac{M_\text{Pl}^2}{M_s^2} = 2\pi(\tau_2')^{2}\mathcal{V}(t') \left( \frac{M_s'}{M_s} \right)^{2} = 2\pi\frac{\tau_2^{2}}{|\tau|^4} \mathcal{V}(t) \left( \frac{M_s'}{M_s} \right)^{8}\,.
\end{equation}
In order for this ratio to remain constant, we must require $\left(
{M_s'}^2/M_s^2 \right) = |\tau|$, or equivalently in terms of the dimensionless
moduli, 
\begin{equation}
	{z'}^a = z^a |\tau|\,,
\end{equation}
which is the transformation used in equation \eqref{Sduality}. Note that when
the axions have vanishing vacuum expectation values as is the case in the main
text, the dilaton takes the simple transformation $\tau_2\to1/\tau_2$.

\paragraph{Fourier--Mukai transform:} in order to reach the D3 limit in cases
where the compact manifold is an elliptic fibration, $X\to B_2$, one needs to
perform two T-dualities, corresponding to an inversion of the dimensionful fibre volume
modulus.
\begin{equation}
	T^1 = \frac{t^1}{M_s^2}\longrightarrow {T''}^1 = \frac{1}{M_s^4 T_1}\,,
\end{equation}
In our case, this corresponds to a particular case of a Fourier--Mukai
transform acting on the dimensionless moduli as dictated by
\eqref{FMTransform}. For simplicity, we will dispense with axion here, as their
vacuum expectation values is set to zero in cases of interest. One finds: 
\begin{equation}\label{FMappendix}
	t^1\longrightarrow \frac{1}{t^1}\,,\qquad 
	t^{\alpha}\longrightarrow  t^\alpha+ \frac{K^\alpha}{2}\left(t^1-\frac{1}{t^1}\right)\,,
\end{equation}
such that 
\begin{align}
	(t^1)''=\frac{1}{t^1}\left(\frac{M_\text{s}''}{M_\text{s}}\right)^2=\frac{1}{t^1}\,.
\end{align}
We used that, conversely to S-duality, these two T-dualities do not change the
string scale: $M_\text{s}''=M_\text{s}$.

From the requirement that the Planck mass should be unchanged under this double
T-duality, we can infer the transformation of the 10d dilaton---the
Buscher's rule---for a non-trivial fibration. To that end, we need the
triple-intersection numbers of the elliptically fibred threefold, which are
related to the intersection pairing of base divisors,
$\eta_{\alpha\beta}=D_\alpha\cdot D_\beta$, and the coefficients of the first
Chern class of the base $c_1(B_2) = K^\alpha D_\alpha$ (see e.g.
\cite{Corvilain:2018lgw}):
\begin{gather}
	\mathcal{K}_{111} = \eta_{\alpha\beta} K^\alpha K^\beta=K^\alpha K_\alpha\,, \qquad  \mathcal{K}_{11\alpha} = \eta_{\alpha\beta} K^\beta= K_\alpha\,,\\
	\mathcal{K}_{1\alpha\beta} = \eta_{\alpha\beta}\,, \qquad  \mathcal{K}_{\alpha\beta\gamma}=0\,,
\end{gather}
where we used the intersection pairing, $\eta_{\alpha\beta}$ to lower indices.
The divisors are chosen as in equations \eqref{verticaldivisor} and
\eqref{sectiondivisor}. The volume is then given by
\begin{equation}
	\mathcal{V} = \frac{1}{6}\mathcal{K}_{abc}t^at^bt^c  =\frac{1}{6}K^\alpha K_\alpha (t^1)^3 + \frac{1}{2} K_\alpha t^\alpha (t^1)^2+\frac{1}{2}t_\alpha t^\alpha t^1\,.
\end{equation}
For completeness, we also give here the volume of these divisors, using
$\text{Vol}(D_a)= \frac{\partial\mathcal{F}}{\partial t^a}$, where $\mathcal{F}$ is the classical prepotential:
\begin{align}
	\text{Vol}(D_1) =& \frac{1}{2}K^\alpha K_\alpha (t^1)^2 + K_\alpha t^\alpha t^1 + \frac{1}{2} t^\alpha t_\alpha\,,\\
	\text{Vol}(D_\alpha) = & t_\alpha t^1 + \frac{1}{2} K_\alpha (t^1)^2\,. 
\end{align}
The volume transforms under the transformation \eqref{FMappendix} as:
\begin{equation}
	\mathcal{V}(t) \longrightarrow \mathcal{V''}(t) = \left(\frac{1}{(t^1)^2}\mathcal{V}(t)+\frac{\eta_{\alpha\beta}K^\alpha K^\beta}{24}((t^1)^{-3}-t^1)\right)\,,
\end{equation}
such that, in a fashion analogous to S-duality, we see that maintaining a
constant Planck-to-string-mass ratio can be achieved by demanding that the 10d
dilaton to transforms as 
\begin{equation}
	\tau_2\longrightarrow \tau_2''=t^1 \tau_2\sqrt{1-\frac{(t^1)^3 \eta_{\alpha\beta}K^\alpha K^\beta}{24\mathcal{V}}-\frac{ \eta_{\alpha\beta}K^\alpha K^\beta}{24\mathcal{V}t^1}}\,.
\end{equation}
The terms proportional to $\eta_{\alpha \beta}K^\alpha K^\beta$ can be
understood as corrections to the usual Buscher's rules due to the non-trivial
elliptic fibration. We hence see that in the limit $t^a \rightarrow \infty$,
corresponding to the D1 limit in the main text, the transformation simplifies 
\begin{align}\label{apptau2transform}
	\tau_2'' \sim t^1\,\tau_2 \,, 
\end{align}
which is the transformation we use for the D3 limit, see equation \eqref{transformtau2}.

%%%%%%%%%%%%%%%%%%%%%%%%%%%%%%%%%%%%%%%%%%%%%%%%%
%%%%%%%%%%%%%%%%%%%%%%%%%%%%%%%%%%%%%%%%%%%%%%%%%
%%%%%%%%%%%%%%%%%%%%%%%%%%%%%%%%%%%%%%%%%%%%%%%%%
\section{Moduli Space Metrics from Contact Potentials}\label{app:metric}
%%%%%%%%%%%%%%%%%%%%%%%%%%%%%%%%%%%%%%%%%%%%%%%%%
%%%%%%%%%%%%%%%%%%%%%%%%%%%%%%%%%%%%%%%%%%%%%%%%%
%%%%%%%%%%%%%%%%%%%%%%%%%%%%%%%%%%%%%%%%%%%%%%%%%
We collect in this appendix some of the expressions needed to evaluate the
moduli space metric subject to quantum corrections in the different limits. In order to know how the line element behaves as one scales the K\"ahler moduli $t^a$ and the 10d dilaton $\tau_2$,
to reach the different limits introduced in section \ref{sec:setup}, one needs
the metric of the hypermultiplet moduli space \eqref{metric}:
\begin{align}
	\mathrm{d}s^2_{\mathcal{M}_\text{HM}} =\frac{1}{2} \left(\mathrm{d}\varphi_4\right)^2 + g_{a \bar{b}} \mathrm{d}z^a \mathrm{d}\bar{z}^b + \text{(axions)} \,.
\end{align} 
While for specific cases the exact hypermultiplet moduli space metric can be calculated explicitly using twistorial methods \cite{Alexandrov:2014sya}, in the limit that we take certain simplifications allow for a simpler approach as shown in appendices C and D of \cite{Marchesano:2019ifh}. In this approach the components of $g_{a \bar b}$ can be obtained---restricting ourselves to
trajectories in the submanifold of $\mathcal{M}_H$ corresponding to vanishing
axion vevs---by taking derivatives of the contact potential $\chi$. In the
D1 limit the contact potential is dominated by D1/D(-1)-instantons, which alter
the classical moduli space metric to:
\begin{eqn}\label{apgab}
    g_{a \bar b}=&-\partial_{z^a}\partial_{\bar z^b} \log \chi \\
      =&\frac{1}{\chi^2}\left[-\frac{\tau_2^2}{8}i\,\mathcal{K}_{ace}t^ct^e-\frac{\tau_2^2}{8\pi} \sum_{k_\Lambda} n_{\bf{k}}^{(0)}  \sum_{m} k_\Lambda \bar z^\Lambda k_aK_0\left(2\pi m |k_\Lambda z^\Lambda| \tau_2\right) \right]\\
      &\times \left[\frac{\tau_2^2}{8}i\,\mathcal{K}_{bdf}t^dt^f-\frac{\tau_2^2}{8\pi} \sum_{k_\Lambda} n_{\bf{k}}^{(0)}  \sum_{m} k_\Lambda z^\Lambda k_b K_0\left(2\pi m |k_\Lambda z^\Lambda| \tau_2\right) \right]\\
     &-\frac{1}{\chi}\left[\frac{\tau_2^2}{8}\mathcal{K}_{abc}t^c-\frac{\tau_2^2}{8\pi} \sum_{k_\Lambda} n_{\bf{k}}^{(0)}  \sum_{m} k_a k_bK_0\left(2\pi m |k_\Lambda z^\Lambda| \tau_2\right)\right.\\
      &\left.+\frac{\tau_2^2}{16\pi}\sum_{k_\Lambda} n_{\bf{k}}^{(0)}  \sum_{m} k_a k_b 2\pi m |k_\Lambda z^\Lambda|\tau_2  K_1\left(2\pi m |k_\Lambda z^\Lambda| \tau_2\right)\right]\, ,
\end{eqn}
where the sums over the instanton corrections are proportional to D1-instanton
charges $k_a$. These terms can be evaluated by performing a Poisson resummation
over $m$, resulting in a sum over a new integer $n$, itself dominated
by the zero mode $n=0$. Up to order one factors, one obtains the following
approximate expression:
\begin{align}
    g_{a \bar b}\supset \frac{\tau_2^2}{\chi }\sum_{ k_\Lambda} n_{\bf{k}}^{(0)}\frac{k_a k_b}{\tau_2|k_\Lambda z^\Lambda|} \,. 
\end{align}
In the region of the asymptotic moduli space along the D1 limit where
D(-1)-instantons are the only significant contribution, \eqref{apgab}
simplifies considerably as we can set the $k_a=0$. The leading contribution to
the metric along the K\"ahler directions is then given by
\begin{align}
    g_{a \bar b}\sim g_{a \bar b}^\text{cl}\frac{\chi^\text{cl}}{\chi}\,,
\end{align}
where $g_{a \bar b}^\text{cl} \sim e^{-2\sigma}$ is the corresponding component
of the classical metric that yields the classical infinite distance. In the
presence of D(-1)-instantons, we have 
\begin{align}
    \chi \gg \chi^\text{cl}
\end{align}
and the metric component along the K\"ahler directions in the region where only
D(-1)-instantons contribute is much smaller than its classical counterpart. 
Following the analysis in \cite{Marchesano:2019ifh}, using the corrected metric to calculate the line element along the K\"ahler directions, their contribution alone would not yield an infinite distance. This turns out also to be true for the quantum corrected metric in the S-dual F1 limit, where we can relate this effect with the fact that these classical coordinates are not the flat ones, and that taking that into account obstructs the limit of vanishing CY volume.

As we saw that the limit we want to take in the K\"ahler direction is
obstructed by quantum corrections such that the initial scaling of the K\"ahler
coordinates is altered, one may now ask what happens to the assumed scaling of
the 10d dilaton in the D1 limit. We can therefore look at the effective metric
components in the direction corresponding to the 10d dilaton that
can be obtained from the metric of the 4d dilaton:
\begin{align}\label{apmetrictau}
    g_{\tau \tau}=\frac{1}{\chi^2}\left(\frac{\partial \chi}{\partial \tau_2}\right)^2 \,,\qquad g_{\tau t^a}=\frac{1}{\chi^2}\left(\frac{\partial \chi}{\partial 
    \tau_2}\right)\left(\frac{\partial \chi}{\partial t^a}\right) \,.
\end{align}
We are mainly interested in the first term for which one finds:
\begin{align}\label{partialchitau}
	\frac{1}{\chi^2}\left(\frac{\partial \chi}{\partial \tau_2}\right)^2=\frac{1}{\chi^2}\left[\frac{\tau_2}{12}\mathcal{K}_{abc}t^at^bt^c-\frac{\tau_2}{8\pi} \sum_{{\bf k_\Lambda}} n_{\bf{k}}^{(0)}  \sum_{m}|k_\Lambda z^\Lambda|^2  K_0\left(2\pi m |k_\Lambda z^\Lambda| \tau_2\right) \right]^2\,.
\end{align}
Note that the sum over the instantons now contains a term proportional to the
D(-1)-instanton charge $k_0$. Thus, for the region of the D1 limit in which
only D(-1)-instantons correct the geometry significantly, the instanton sum does
not vanish in contrast to what happened to $g_{a \bar b}$ in the same region.
Poisson resumming the instanton sum, we obtain the leading term for $g_{\tau
\tau}$:
\begin{align}
     \frac{1}{\chi^2}\left(\frac{\partial \chi}{\partial \tau_2}\right)^2 \sim \frac{1}{\chi^2}\left[\frac{1}{16 \pi}\sum_{{\bf k_\Lambda}} n_{\bf{k}}^{(0)} |k_\Lambda z^\Lambda|\right]^2\,. 
\end{align}
Evaluating the sum over $k_0$ we obtain 
\begin{align}
    g_{\tau \tau}\simeq \frac{1}{\chi^2} \left[\frac{\chi_E(X)}{16\pi}\left(k_0^\text{max}\right)^2\right]^2\sim \left(k_0^\text{max}\right)^2\,,
\end{align}
which should be compared with the classical scaling,
\begin{align}
    g_{\tau \tau}^\text{cl}\simeq e^{3\sigma}\,. 
\end{align}
We thus see that the metric in the $\tau_2$ direction receives strong quantum
corrections but since $k_0^\text{max}\sim e^{3/2 \sigma}$ the scaling of the
classical metric is recovered. The scaling of the 10d dilaton---interpreted as
a coordinate of the moduli space---therefore does not receive any quantum corrections
itself. In section \ref{sec:QuantumCorrections}, we use this fact to infer that
the scaling of the tension of a D1-string in string units is not affected by
quantum corrections.

%%%%%%%%%%%%%%%%%%%%%%%%%%%%%%%%%%%%%%%%%%%%%%%%%
%%%%%%%%%%%%%%%%%%%%%%%%%%%%%%%%%%%%%%%%%%%%%%%%%
%%%%%%%%%%%%%%%%%%%%%%%%%%%%%%%%%%%%%%%%%%%%%%%%%
\section{Contact Potential for Mutually Local Instantons}\label{app:D3-instantons}
%%%%%%%%%%%%%%%%%%%%%%%%%%%%%%%%%%%%%%%%%%%%%%%%%
%%%%%%%%%%%%%%%%%%%%%%%%%%%%%%%%%%%%%%%%%%%%%%%%%
%%%%%%%%%%%%%%%%%%%%%%%%%%%%%%%%%%%%%%%%%%%%%%%%%

As shown in section \ref{sec:QuantumCorrections}, D1/D(-1)-instantons are
mapped to D3/D1/D(-1)-instantons under a Fourier--Mukai transform. We show
here how they can be incorporated to the quantum corrected 4d dilaton. 

For generic instanton effects (excluding NS5-instantons) and with vanishing
axion vacuum expectation values, these corrections are shown to be given by
\cite{Alexandrov:2011va,Alexandrov:2013yva}:
\begin{align}
    e^{-2\varphi_4}=&\frac{1}{12}\tau_2^2 \mathcal{K}_{abc}t^at^bt^c -\frac{\chi_E(X)}{192 \pi}\nonumber\\
    &-\frac{i\tau_2}{64\pi^2}e^{K/2}\sum_\gamma \Omega(\gamma)\int_{l_\gamma} \frac{dt}{t}\left(t^{-1}Z_\gamma - t \bar Z_\gamma \right)\log\left(1-\lambda_D(\gamma)\mathcal{X}_\gamma\right)\,.
\end{align}
The so-far-undefined quantities denote the following: the Donaldson--Thomas
invariants $\Omega(\gamma)$, and the central charge $Z_\gamma$ associated to
the instanton charge vector $\gamma$; the integral is performed along a BPS ray
$l_\gamma$; the integrand depends on a Fourier mode $\mathcal{X}_\gamma$.  In
cases where the instantons of interest are mutually local, which is the case
here as D3/D1/D(-1)-instantons are obtained by the monodromy transformation
associated to the Fourier--Mukai transform, $\mathcal{X}_\gamma$ reduces
to the so-called ``semi-flat'' Fourier mode:
\begin{align}
    \mathcal{X}_\gamma^\text{sf}(t)=\exp\left[-\pi i\tau_2e^{-\mathcal{K}/2}\left(t^{-1}Z_\gamma -t\bar Z_\gamma\right)\right]\,.
\end{align}
Using a Taylor expansion for the logarithm around $x=1$, the integral then
reduces to a Bessel function, such that the quantum corrections to the
4d dilaton associated to D3/D1/D(-1)-instantons are given by:
\begin{align}
    e^{-2\varphi} =\frac{1}{12}\tau_2^2 \mathcal{K}_{abc}t^at^bt^c -\frac{\chi(X)}{192 \pi}+\frac{\tau_2}{8\pi^2}\sum_\gamma \Omega(\gamma)\sum_{m=1}^\infty \frac{\text{Re}(S_\gamma)}{m}K_1(2\pi m \text{Re}(S_\gamma))\,,
\end{align}
where the instanton action $S_\gamma$ has been found in equation
\eqref{instantonactionD3}. One can see that, as expected, it takes the form
similar to the correction associated to the original D1/D(-1)-instantons, which
as explained in section \ref{sec:QuantumCorrections} is related to the fact
that the Fourier--Mukai transform can be interpreted as a rotation in the space
of charges.

%\printbibliography
%%%%%%%%%%%%%%%%%%%%%%%%%%%%%%%%%%%%%%%%%%%%%%%%%
%%%%%%%%%%%%%%%%%%%%%%%%%%%%%%%%%%%%%%%%%%%%%%%%%
%%%%%%%%%%%%%%%%%%%%%%%%%%%%%%%%%%%%%%%%%%%%%%%%%

\bibliography{references}{}
\bibliographystyle{JHEP} 

%%%%%%%%%%%%%%%%%%%%%%%%%%%%%%%%%%%%%%%%%%%%%%%%%
%%%%%%%%%%%%%%%%%%%%%%%%%%%%%%%%%%%%%%%%%%%%%%%%%
%%%%%%%%%%%%%%%%%%%%%%%%%%%%%%%%%%%%%%%%%%%%%%%%%
\end{document}